\documentclass[twocolumn,showpacs,aps,prd,floatfix]{revtex4}
\usepackage{epsfig}
\usepackage{rotating}
\usepackage{multirow}
\usepackage{morefloats}

\RequirePackage{xspace}

\newcommand{\DBR}{\mbox {$\Delta \ensuremath{\cal B}$}}
\def\smaller{\footnotesize}
\def\babar{\mbox{\slshape B\kern-0.1em{\smaller A}\kern-0.1em
    B\kern-0.1em{\smaller A\kern-0.2em R}}}
\def\invfb   {\ensuremath{\mbox{\,fb}^{-1}}\xspace}
\def\Y#1S{\ensuremath{\Upsilon{(#1S)}}\xspace}
\def\FourS {\Y4S}
\def\jpsi     {\ensuremath{{J\mskip -3mu/\mskip -2mu\psi\mskip 2mu}}\xspace}
\def\pep2{PEP-II}
\def\mev {\ensuremath{\mathrm{\,Me\kern -0.1em V}}\xspace}
\def\gev {\ensuremath{\mathrm{\,Ge\kern -0.1em V}}\xspace}
\def\gevcc {\ensuremath{{\mathrm{\,Ge\kern -0.1em V\!/}c^2}}\xspace}
\def\mes        {\mbox{$m_{ES}$}\xspace}
\def\DeltaE     {\mbox{$\Delta E$}\xspace}
\def\KL    {\ensuremath{K^0_{\scriptscriptstyle L}}\xspace} 
\def\BR {\ensuremath{\cal B}\xspace}

\def\ulnu{$b \rightarrow u \ell \nu$}
\def\clnu{$b \rightarrow c \ell \nu$}

\def\pilnu{$B^{0} \rightarrow \pi^{-} \ell^{+} \nu$}
\def\etalnu{$B^{+} \rightarrow \eta \ell^{+} \nu$}
\def\etaplnu{$B^{+} \rightarrow \eta^{\prime} \ell^{+} \nu$}
\def\omegalnu{$B^+ \rightarrow \omega \ell^+ \nu$}
\def\etaetaplnu{$B^{+} \rightarrow \eta^{(\prime)} \ell^{+} \nu$}
\def\piclnu{$B \rightarrow \pi \ell^+ \nu$}
\def\pizlnu{$B^{+} \rightarrow \pi^{0} \ell^{+} \nu$}
\def\rholnu{$B \rightarrow \rho \ell \nu$}
\def\dlnu{$B \rightarrow D \ell \nu$}
\def\dstrlnu{$B \rightarrow D^* \ell \nu$}
\def\ddblstrlnu{$B \rightarrow D^{**} \ell \nu$}
\def\bfpiclnu{${\BR}(B \rightarrow \pi \ell^{+} \nu)$}
\def\bfpilnu{${\BR}(B^{0} \rightarrow \pi^{-} \ell^{+} \nu)$}
\def\bfpiZlnu{${\BR}(B^{+} \rightarrow \pi^{0} \ell^{+} \nu)$}
\def\bfetalnu{${\BR}(B^{+} \rightarrow \eta \ell^{+} \nu)$}
\def\bfetaplnu{${\BR}(B^{+} \rightarrow \eta^{\prime} \ell^{+} \nu)$}
\def\Bz      {\ensuremath{B^0}\xspace}
\def\Bu      {\ensuremath{B^+}\xspace}
\def\Bbar    {\kern 0.18em\overline{\kern -0.18em B}{}\xspace}
\def\Bzb     {\ensuremath{\Bbar^0}\xspace}
\def\Bub     {\ensuremath{B^-}\xspace}
\def\BzBzb   {\ensuremath{\Bz {\kern -0.16em \Bzb}}\xspace}
\def\BpBm    {\ensuremath{\Bu {\kern -0.16em \Bub}}\xspace}

\def\bfupsbzbz{${\BR}(\Upsilon(4S) \rightarrow B^0\bar{B^0})$}

\def\bfupsbpbm{${\BR}(\Upsilon(4S) \rightarrow B^+B^-)$}
\def\bfrhoClnu{${\BR}(B^{0} \rightarrow \rho^{-} \ell^{+} \nu)$}
\def\bfrhoZlnu{${\BR}(B^{+} \rightarrow \rho^{0} \ell^{+} \nu)$}
\def\bfomegalnu{${\BR}(B^{+} \rightarrow \omega \ell^{+} \nu)$}
\def\bfDlnu{${\BR}(B \rightarrow D \ell \nu)$}
\def\bfDstrlnu{${\BR}(B \rightarrow D^* \ell \nu)$}
\def\bfDdblstrlnu{${\BR}(B \rightarrow D^{**} \ell \nu)$}
\def\bfetaetaplnu{${\BR}(B^{+} \rightarrow \eta^{(\prime)} \ell^{+} \nu)$}
\def\bfpiclnuq{$\Delta {\BR}(B \rightarrow \pi \ell^{+} \nu,q^2)$}
\def\bfpizlnuq{$\Delta {\BR}(B^{+} \rightarrow \pi^{0} \ell^{+} \nu,q^2)$}
\def\bfpilnuq{$\Delta {\BR}(B^{0} \rightarrow \pi^{-} \ell^{+} \nu,q^2)$}
\def\bfomegalnuq{$\Delta {\BR}(B^+ \rightarrow \omega \ell^{+} \nu,q^2)$}
\def\bfetalnuq{$\Delta {\BR}(B^{+} \rightarrow \eta \ell^{+} \nu,q^2)$}

\def\bfpilnuqq{$\DBR(q^2)$}

\def\lnrt{loose neutrino reconstruction technique}

\def\qq{$q^2$}
\def\qqr{$q^2$} 
\def\fplus{$f_+(q^2)$}
\def\vub{$|V_{ub}|$}

\def\udsctau{$u\bar{u}/d\bar{d}/s\bar{s}/c\bar{c}/\tau^+\tau^-$}

\def\obb{other $B\bar{B}$}

\def\DeltaEDef{\ensuremath{\Delta E = (P_B \cdot P_{beams} - s/2) / \sqrt{s}}
\xspace}
\def\FitRegion{\ensuremath{|\Delta E| < 1.0~\gev \mbox{ and } m_{ES} > 
5.19~\gev}\xspace}
\newcommand{\gevsq}{\ensuremath{\mathrm{\,Ge\kern -0.1em V^2\!}}\xspace}

\def\VubFpz{$|V_{ub}f_+(0)|$}
\def\VubFpzVala{$8.7 \pm 0.4$}
\def\VubFpzValza{$9.1 \pm 0.5$}
\def\VubFpzValca{$8.7 \pm 0.3$}
\def\VubFpzValb{$8.6 \pm 0.5$}
\def\VubFpzValzb{$9.4 \pm 0.6$}
\def\VubFpzValcb{$8.8 \pm 0.4$}

\newcommand{\BABARPubYear}{12}
\newcommand{\BABARPubNumber}{015}
\newcommand{\SLACPubNumber}{15208}

\begin{document}

\preprint{\babar-PUB-\BABARPubYear/\BABARPubNumber} 
\preprint{SLAC-PUB-\SLACPubNumber} 

\begin{flushleft}
 \babar-PUB-\BABARPubYear/\BABARPubNumber\\
 SLAC-PUB-\SLACPubNumber\\
arXiv:1208.1253 [hep-ex] \\
\end{flushleft}

\title{\large\bf\boldmath
Branching fraction and form-factor shape measurements of exclusive charmless  
semileptonic $B$ decays, and determination of \vub}

%
\author{J.~P.~Lees}
\author{V.~Poireau}
\author{V.~Tisserand}
\affiliation{Laboratoire d'Annecy-le-Vieux de Physique des Particules (LAPP), Universit\'e de Savoie, CNRS/IN2P3,  F-74941 Annecy-Le-Vieux, France}
\author{J.~Garra~Tico}
\author{E.~Grauges}
\affiliation{Universitat de Barcelona, Facultat de Fisica, Departament ECM, E-08028 Barcelona, Spain }
\author{A.~Palano$^{ab}$ }
\affiliation{INFN Sezione di Bari$^{a}$; Dipartimento di Fisica, Universit\`a di Bari$^{b}$, I-70126 Bari, Italy }
\author{G.~Eigen}
\author{B.~Stugu}
\affiliation{University of Bergen, Institute of Physics, N-5007 Bergen, Norway }
\author{D.~N.~Brown}
\author{L.~T.~Kerth}
\author{Yu.~G.~Kolomensky}
\author{G.~Lynch}
\affiliation{Lawrence Berkeley National Laboratory and University of California, Berkeley, California 94720, USA }
% ...
\collaboration{The \babar\ Collaboration}
\noaffiliation

\begin{abstract}
 We report the results of a study of the exclusive charmless semileptonic 
decays, $B^{0} \rightarrow \pi^{-} \ell^{+} \nu$, $B^{+} \rightarrow \pi^{0} 
\ell^{+} \nu$, $B^{+} \rightarrow \omega \ell^{+} \nu$, $B^{+} \rightarrow \eta
\ell^{+} \nu$ and $B^{+} \rightarrow \eta^{\prime} \ell^{+} \nu$, ($\ell = e$ 
or $\mu$) undertaken with approximately $462\times 10^6$ $B\bar{B}$ pairs 
collected at the $\Upsilon(4S)$ resonance with the 
${\mbox{\slshape B\kern-0.1em{\smaller A}\kern-0.1em 
B\kern-0.1em{\smaller A\kern-0.2em R}}}$ detector. The analysis uses events in 
which the signal $B$ decays are reconstructed with a loose neutrino 
reconstruction technique. We obtain partial branching fractions in several bins
of $q^2$, the square of the momentum transferred to the lepton-neutrino pair, 
for $B^0 \rightarrow \pi^-\ell^+\nu$, $B^+ \rightarrow \pi^0\ell^+ \nu$, $B^{+}
\rightarrow \omega \ell^{+} \nu$ and $B^{+} \rightarrow \eta \ell^{+} \nu$. 
From these distributions, we extract the  form-factor shapes $f_+(q^2)$ and the
total branching fractions ${\ensuremath{\cal B}}(B^0 \rightarrow \pi^- \ell^+ 
\nu)$ $ = \left(1.45 \pm 0.04_{stat} \pm 0.06_{syst} \right) \times 10^{-4}$ 
(combined $\pi^-$ and $\pi^0$ decay channels assuming isospin symmetry), 
${\ensuremath{\cal B}}(B^{+} \rightarrow \omega \ell^{+} \nu)$ $ = \left(1.19 
\pm 0.16_{stat} \pm 0.09_{syst} \right) \times 10^{-4}$ and 
${\ensuremath{\cal B}}(B^{+} \rightarrow \eta \ell^{+} \nu)$ $ = \left(0.38 
\pm 0.05_{stat} \pm 0.05_{syst} \right) \times 10^{-4}$. We also measure 
${\ensuremath{\cal B}}(B^{+} \rightarrow \eta^{\prime} \ell^{+} \nu)$ $ = 
\left(0.24 \pm 0.08_{stat} \pm 0.03_{syst} \right) \times 10^{-4}$. We obtain 
values for the magnitude of the CKM matrix element $\ensuremath{|V_{ub}|}$ by 
direct comparison with three different QCD calculations in restricted $q^2$ 
ranges of $B \rightarrow \pi\ell^+\nu$ decays. From a simultaneous fit to the 
experimental data over the full $q^2$ range and the FNAL/MILC lattice QCD 
predictions, we obtain $\ensuremath{|V_{ub}|} = (3.25 \pm 0.31)\times 10^{-3}$,
where the error is the combined experimental and theoretical uncertainty.  
\end{abstract}

\pacs{13.20.He,                 
      12.15.Hh,                 
      12.38.Qk,                 
      14.40.Nd}                 

\maketitle  
\parskip=0.0cm
\abovecaptionskip=0.0cm
\belowcaptionskip=0.0cm

\section{Introduction}

 A precise measurement of the CKM matrix~\cite{CKM} element \vub\ will 
improve our quantitative understanding of weak interactions and CP violation
in the Standard Model. The value of \vub\ can be determined by the measurement 
of the partial branching fractions of exclusive charmless semileptonic $B$ 
decays since the rate for decays that involve a scalar meson is 
proportional to $|V_{ub}f_+(q^2)|^2$. Here, the form factor \fplus\ depends on 
\qq, the square of the momentum transferred to the lepton-neutrino pair. Values
of \fplus\ can be calculated at small \qq\ ($\lesssim$ 16 \gevsq) using Light 
Cone Sum Rules (LCSR)~\cite{LCSR, LCSR2, singlet} and at large \qq\ ($\gtrsim$ 
16 \gevsq) from unquenched Lattice QCD (LQCD)~\cite{HPQCD06,FNAL}. Extraction 
of the \fplus\ form-factor shapes from exclusive decays~\cite{PlusCC} such as 
\pilnu~\cite{Jochen, Simard, Belle}, \pizlnu~\cite{Jochen}, 
\omegalnu~\cite{Wulsin} and \etaetaplnu~\cite{Simard} may be used to test these
theoretical predictions~\cite{PDG10}. Measurements of the branching fractions 
(BF) of all these decays will also improve our knowledge of the composition of 
charmless semileptonic decays. This input can be used to reduce the large 
systematic uncertainty in $|V_{ub}|$ due to the poorly known \ulnu\ signal 
composition in inclusive semileptonic $B$ decays. It will also help to 
constrain the size of the gluonic singlet contribution to form factors for the 
\etaetaplnu\ decays~\cite{singlet, singlet2}. 

 In this paper, we present measurements of the partial BFs \bfpilnuq\ in 12 
bins of \qq, \bfpizlnuq\ in 11 bins of \qq, \bfomegalnuq\ and \bfetalnuq\ in 
five bins of \qq, as well as the BF \bfetaplnu. From these distributions, we 
extract the total BFs for each of the five decay modes. Values of these BFs 
were previously reported in Refs.~\cite{Jochen, Simard, Belle, Wulsin}, and 
references therein. In this work, we carry out an untagged analysis ({\it i.e.}
the second $B$ meson is not explicitly reconstructed) with the 
\lnrt~\cite{Cote} whereby the selections on the variables required to 
reconstruct the neutrino are much looser than usual. This results in a large 
candidate sample. Concerning the \pizlnu\ and \omegalnu\ decay modes, this is 
the first analysis using this technique.
 
 We assume isospin symmetry to hold, and combine the data of the \pizlnu\ and 
\pilnu\ channels thereby leading to a large increase, of the order of 34\%, in 
the effective number of \pilnu\ events available for study. We refer to such 
events as \piclnu\ decays. The values of the BFs obtained in the present work 
are based on the use of the most recent BFs and form-factor shapes for all 
decay channels in our study. In particular, the subsequent improved 
treatment of the distributions that describe the combination of resonant and 
nonresonant \ulnu\ decays results in an increase of 3.5\% in the total BF value
of the \pilnu\ decays. This increase is significant in view of the total 
uncertainty of 5.1\% obtained in the measurement of this BF. 
 
 We now optimize our selections over the entire fit region instead of the 
signal-enhanced region, as was done previously~\cite{Simard}. The ensuing  
tighter selections produce a data set with a better signal to background ratio 
and higher purity in the \pilnu\ and \etaetaplnu\ decays. As a result, we can 
now investigate the \etaetaplnu\ decays over their full \qq\ ranges. The 
present analysis of the \pilnu\ decay channel makes use of the full \babar\ 
data set compared to only a subset in Ref.~\cite{Jochen}. As for the \omegalnu\
decay channel, it uses the unfolded values of the partial BFs and a selection
procedure that is significantly different from the one in Ref.~\cite{Wulsin}.
The unfolding process is used to obtain the distribution of the true values of 
$q^2$ by applying the inverse of the detector response matrix to the 
distribution of the measured values of $q^2$. Each element of this matrix is 
constructed in MC simulation for each bin of \qqr\ as the ratio of the number 
of true events to the total number of reconstructed events. The current work 
provides results for five decay channels using the same analysis method.
   
 In this work, we compare the values of \bfpilnuqq\ for the \piclnu\ mode 
to form-factor calculations \cite{LCSR, LCSR2, HPQCD06, FNAL} in restricted 
\qq\ ranges to obtain values of \vub. Values of \vub\ with a smaller total 
uncertainty can also be obtained from a simultaneous fit to the \piclnu\ 
experimental data over the full \qq\ range and the FNAL/MILC lattice QCD 
predictions~\cite{FNAL}. Such values were recently obtained by 
\babar~\cite{Jochen} ($|V_{ub}|=(2.95\pm0.31)\times10^{-3}$) and 
Belle~\cite{Belle} ($|V_{ub}|=(3.43\pm0.33)\times10^{-3}$). These results are 
consistent at the 2$\sigma$ level, when taking into account the correlations,
but display a tension with respect to the value of \vub\ measured~\cite{PDG10} 
in inclusive semileptonic $B$ decays, $|V_{ub}|=(4.27\pm0.38)\times10^{-3}$. 
This study attempts to resolve the tension by analyzing the data using the most
recent values of BFs and form factors.

\section{Data Sample and Simulation}

 We use a sample of $462\times 10^6$ $B\bar{B}$ pairs, corresponding to an 
integrated luminosity of 416.1~\invfb, collected at the \FourS\ resonance with 
the \babar\ detector~\cite{ref:babar} at the \pep2\ asymmetric-energy $e^+e^-$ 
storage rings. A sample of 43.9~\invfb\ collected approximately 40 \mev\ 
below the \FourS\ resonance (denoted ``off-resonance data'') is used to study 
contributions from $e^+e^- \rightarrow$ \udsctau\ (continuum) events. Detailed 
Monte Carlo (MC) simulations are used to optimize the signal selections, 
estimate the signal efficiencies, obtain the shapes of the signal and 
background distributions and determine the systematic uncertainties 
associated with the BF values. 

 MC samples are generated for $\Upsilon (4S) \rightarrow B\bar{B}$ events, 
continuum events, and dedicated signal samples containing \pilnu, \pizlnu, 
\omegalnu\ and \etaetaplnu\ signal decays, separately. These signal MC events 
are produced with the FLATQ2 generator~\cite{bad809}. The $f_+(q^2)$ shape
used in this generator is adjusted by reweighting the generated events. For the
\piclnu\ decays, the signal MC events are reweighted to reproduce the 
Boyd-Grinstein-Lebed (BGL) parametrization~\cite{BGL}, where the parameters are
taken from Ref.~\cite{Simard}. For the \omegalnu\ decays, the events are 
reweighted to reproduce the Ball parametrization~\cite{Ball05}. For the 
\etaetaplnu\ decays, the signal MC events are reweighted to reproduce the 
Becirevic-Kaidalov (BK) parametrization~\cite{BK}, where the parameter 
$\alpha_{BK} = 0.52\pm0.04$ gave a reasonable fit to the \pilnu\ and \etalnu\ 
data of Ref.~\cite{Simard}. The \babar\ detector's acceptance and response are 
simulated using the GEANT4 package~\cite{Geant4}.

\section{Event Reconstruction and Candidate Selection}

 To reconstruct the decays \pilnu, \pizlnu, \omegalnu\ and \etaetaplnu, we 
first reconstruct the final state meson. The $\omega$ meson is reconstructed in
the $\omega \rightarrow \pi^+\pi^-\pi^0$ decay channel. The $\eta$ meson is 
reconstructed in the $\eta \rightarrow \gamma\gamma$ ($\eta(\gamma\gamma)$) and
$\eta \rightarrow \pi^+\pi^-\pi^0$ ($\eta(3\pi)$) decay channels while the 
$\eta^{\prime}$ is reconstructed in the $\eta^{\prime} \rightarrow 
\eta\pi^+\pi^-$ decay channel, followed by the $\eta \rightarrow \gamma\gamma$ 
decay ($\eta^{\prime}(\gamma\gamma)$). The $\eta^{\prime} \rightarrow \rho^0 
\gamma$ decay channel suffers from large backgrounds and we do not consider it 
in the present work. 

 Event reconstruction with the \babar\ detector is described in detail 
elsewhere~\cite{ref:babar}. Electrons and muons are mainly identified by their 
characteristic signatures in the electromagnetic calorimeter and the 
muon detector, respectively, while charged hadrons are identified and 
reconstructed using the silicon vertex tracker, the drift chamber and the
Cherenkov detector. The photon and charged particle tracking reconstruction
efficiencies are corrected using various control samples. The average 
electron and muon reconstruction efficiencies are 93\% and 70\%, respectively, 
while the corresponding probabilities that a pion is identified as a lepton are
less than $0.2$\% and less than $1.5$\%, respectively. 

  The neutrino four-momentum, $P_{\nu}=(|\vec{p}^{_*}_{miss}|,
\vec{p}^{_*}_{miss})$, is inferred from the difference between the momentum of 
the colliding-beam particles $\vec{p}^{_*}_{beams}$ and the vector sum of the 
momenta of all the particles detected in the event $\vec{p}^{_*}_{tot}$, such 
that $\vec{p}^{_*}_{miss}=\vec{p}^{_*}_{beams}-\vec{p}^{_*}_{tot}$. All 
variables with an asterisk are given in the \FourS\ frame. To evaluate 
$E_{tot}$, the total energy of all detected particles, we assume zero mass for 
all neutral candidates, and we use the known masses for the charged particles 
identified in the event. If the particle is not identified, its mass is assumed
to be that of a pion.  

  In this analysis, we calculate the momentum transfer squared as 
$q^2=(P_{B}-P_{meson})^2$ instead of $q^2=(P_{\ell}+P_{\nu})^2$, where $P_B$, 
$P_{meson}$ and $P_{\ell}$ are the four-momenta of the $B$ meson, of the $\pi$,
$\omega$, $\eta$ or $\eta^{\prime}$ meson, and of the lepton, respectively, 
evaluated in the \FourS\ frame. With this choice, the value of \qqr\ is 
unaffected by any misreconstruction of the neutrino. To maintain this 
advantage, $P_B$ must be evaluated without any reference to the neutrino. It 
has an effective value since the magnitude of the 3-momentum $\vec{p}^{_*}_B$ 
is determined from the center-of-mass energy and the known $B$ meson mass but 
the direction of the $B$ meson cannot be measured. It can only be estimated. 

  To do this, we first combine the lepton with a 
$\pi$, $\omega$, $\eta$ or $\eta^{\prime}$ meson to form the so-called $Y$ 
pseudoparticle such that $P_Y = P_{\ell}+P_{meson}$. The angle $\theta_{BY}$, 
between the $Y$ and $B$ momenta in the \FourS\ frame, can be determined under 
the assumption that the only unobserved decay product is a neutrino, 
{\it i.e.}, $B \rightarrow Y\nu$. In this frame, the $Y$ momentum, the $B$ 
momentum and the angle $\theta_{BY}$ define a cone with the $Y$ momentum as its
axis and with a true $B$ momentum lying somewhere on the surface of the cone. 
The $B$ rest frame is thus known up to an azimuthal angle $\psi$ about the $Y$ 
momentum. The value of \qqr\ is then computed, as explained in 
Ref.~\cite{DstrFF}, as the average of four \qqr\ values corresponding to four 
possible angles, $\psi$, $\psi+\pi/2$, $\psi+\pi$, $\psi+3\pi/2$ rad, where the
angle $\psi$ is chosen randomly. The four values of \qqr\ are weighted by 
the factor $\sin^2\theta_B$, $\theta_B$ being the angle between the $B$ 
direction and the beam direction in the \FourS\ frame. This weight is needed 
since $B\bar{B}$ production follows a $\sin^2\theta_B$ distribution in the 
\FourS\ frame. We require that $|\cos\theta_{BY}| \leq 1$. We correct for the 
reconstruction effects on the measured values of \qq\ (the \qq\ resolution is
approximately 0.6 \gevsq) by applying an unregularized unfolding algorithm to 
the measured \qq\ spectra~\cite{Cowan}. 

 The selections of the candidate events are determined in MC simulation by 
maximizing the ratio $S/\sqrt{(S+B)}$ over the entire fit region, where $S$ is 
the number of correctly reconstructed signal events and $B$ is the total number
of background events. The continuum background is suppressed by requiring the 
ratio of second to zeroth Fox-Wolfram moments~\cite{FW} to be smaller than 0.5.
Radiative Bhabha and two-photon processes are rejected by requirements on the 
number of charged particle tracks and neutral calorimeter 
clusters~\cite{BhabhaVeto}. To ensure all track momenta are well measured, 
their polar angles are required to lie between 0.41 and 2.46 rad with respect 
to the electron beam direction (the acceptance of the detector). For all 
decays, we demand the momenta of the lepton and meson candidates to be 
topologically compatible with a real signal decay by requiring that a 
mass-constrained geometrical vertex fit~\cite{mass} of the tracks associated 
with the two particles gives a $\chi^{2}$ probability greater than 1\%. In the 
fit, the external constraints such as reconstructed tracks are treated first,
followed by all four-momenta conservation constraints. Finally, at each
vertex, the geometric constraints and the mass are combined. These combined
constraints are applied consecutively.

  To reduce the number of unwanted leptons and secondary decays such as 
$D \rightarrow X\ell\nu$, \jpsi, $\tau$ and kaon decays, the minimum transverse
momentum is 50 \mev\ for all leptons and 30 \mev\ for all photons, and 
all electron (muon) tracks are required to have momenta greater than 0.5 (1.0) 
\gev\ in the laboratory frame. The momenta of the lepton and the meson are 
further restricted to enhance signal over background. We require the following:
\begin{itemize}
 \item for \piclnu\ decays: \\
  $|\vec{p}^{_*}_{\ell}|>2.2$ \gev\ or $|\vec{p}^{_*}_{\pi}|>1.3$ \gev\ \\
or $|\vec{p}^{_*}_{\ell}|+|\vec{p}^{_*}_{\pi}|>2.8$ \gev; 
 \item for \omegalnu\ decays: \\ 
$|\vec{p}^{_*}_{\ell}|>2.0$ \gev\ or $|\vec{p}^{_*}_{\omega}|>1.3$ \gev\ \\
or $|\vec{p}^{_*}_{\ell}|+|\vec{p}^{_*}_{\omega}|>2.65$ \gev; 
 \item for \etalnu\ decays: \\
$|\vec{p}^{_*}_{\ell}|>2.1$ \gev\ or $|\vec{p}^{_*}_{\eta}|>1.3$ \gev\ \\ 
or $|\vec{p}^{_*}_{\ell}|+|\vec{p}^{_*}_{\eta}|>2.8$ \gev;
 \item for \etaplnu\ decays: \\
 $|\vec{p}^{_*}_{\ell}|>2.0$ \gev\ or $|\vec{p}^{_*}_{\eta^{\prime}}|>1.65$ 
\gev\ \\
or $0.69|\vec{p}^{_*}_{\ell}|+|\vec{p}^{_*}_{\eta^{\prime}}|>2.4$ \gev.
\end{itemize} 
These cuts primarily reject background and reduce the signal efficiencies by 
less than 5\%. 

   To remove $\jpsi \rightarrow \mu^{+} \mu^{-}$ decays, we reject any 
combination of two muons, including misidentified pions, if the two particles  
have an invariant mass consistent with the \jpsi\ mass [3.07-3.13] \gev. We do 
not apply a specific \jpsi\ veto for $\jpsi \rightarrow e^+e^-$ decays, since 
we find no evidence for any remaining such events in our data set. We restrict 
the reconstructed masses of the meson to lie in the interval:
\begin{itemize}
 \item for \pizlnu\ decays: $0.115<m_{\pi^0}<0.150$ \gev,
 \item for \omegalnu\ decays: $0.760<m_{\omega}<0.805$ \gev, 
 \item for \etalnu\ decays: $0.51<m_{\eta}<0.57$ \gev,
 \item for \etaplnu\ decays: $0.92<m_{\eta^{\prime}}<0.98$ \gev. 
\end{itemize}

 Backgrounds are further reduced by \qqr-dependent selections on the cosine of 
the angle, $\cos\theta_{thrust}$, between the thrust axes~\cite{thrust} of the 
$Y$ and of the rest of the event; on the polar angle, $\theta_{miss}$, 
associated with $\vec{p}_{miss}$; on the invariant missing mass squared, 
$m^2_{miss}=E^2_{miss}- |\vec{p}_{miss}|^2$, divided by twice the missing 
energy ($E_{miss}= E_{beams} - E_{tot}$); on the cosine of the angle,
 $\cos\theta_{\ell}$, between the direction of the virtual $W$ boson ($\ell$ 
and $\nu$ combined) boosted in the rest frame of the $B$ meson and the 
direction of the lepton boosted in the rest frame of the $W$ boson; and on L2, 
the momentum weighted Legendre monomial of order 2. The quantity  
$m^2_{miss}/2E_{miss}$ should be consistent with zero if a single neutrino is 
missing. The phrase ``rest of the event'' refers to all the particles left in 
the event after the lepton and the meson used to form the Y pseudoparticle are 
removed. 

\begin{figure}
\begin{center}
\epsfig{file=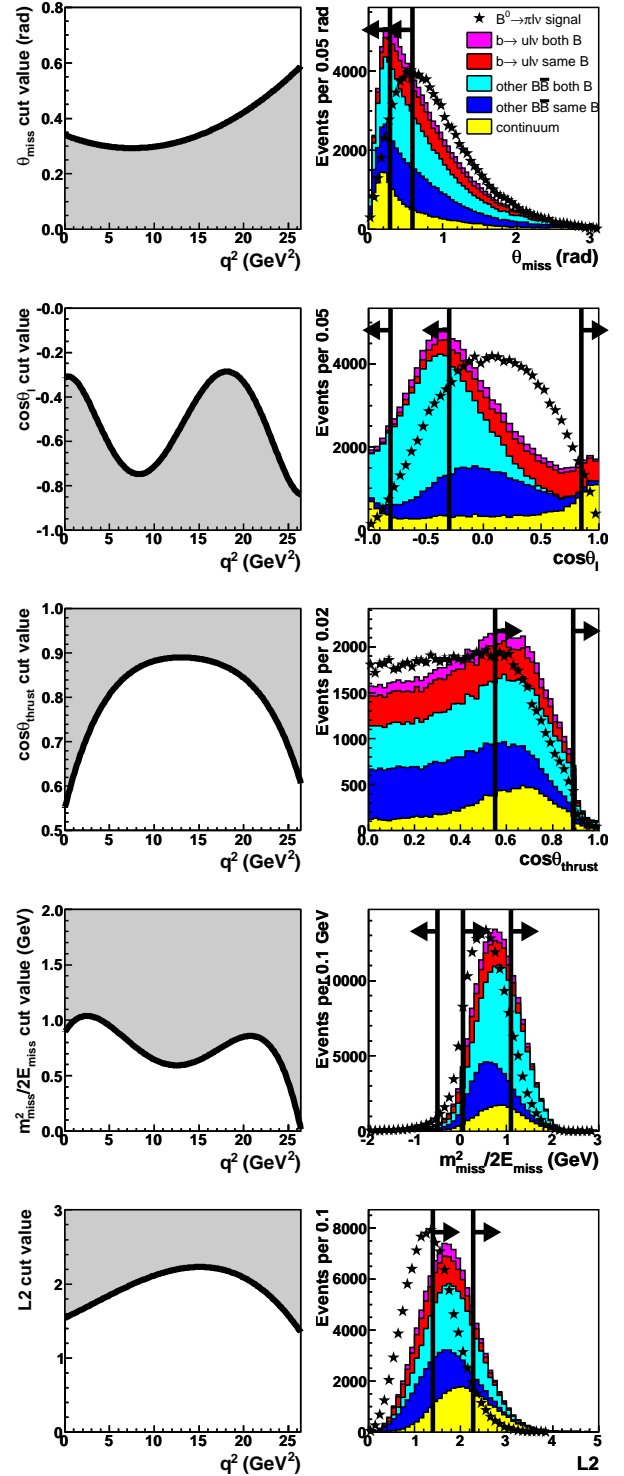,height=20.0cm}
\caption[]{\label{CUTJPG} (color online) Left panels: Distributions of the 
selection values for the $q^2$-dependent selections on the variables used in 
the analysis of \pilnu\ decays. The vertical axis represents the selection 
value for a given \qqr\ value. We reject an event when its value is in the 
shaded region. Right panels: Corresponding distributions in the total fit 
region illustrating the effects of the $q^2$-dependent selections. The arrows 
indicate the rejected regions, as explained in the text. All the selections 
have been applied except for the one of interest. In each panel, the signal 
area is scaled to the area of the total background.}
\end{center}
\end{figure}

 The \qq-dependent selections are shown in the panels on the left-hand side of 
Fig.~\ref{CUTJPG}, and their effects are illustrated in the panels on the 
right-hand side of the same figure, for \pilnu\ decays. A single vertical line 
indicates a fixed cut, a set of two vertical lines represents a \qq-dependent 
cut. The position of the two lines corresponds to the minimum and maximum 
values of the selection, shown in the left-hand side panels. The functions 
describing the \qq\ dependence are given in 
Tables~\ref{cutSummaryPic}-\ref{cutSummary} of the Appendix for the 
five decays under study. For \etalnu\ decays, additional background is rejected
by requiring that $|\cos\theta_{V}|<0.95$, where $\theta_{V}$ is the helicity 
angle of the $\eta$ meson~\cite{bad809}. 

 The kinematic variables \DeltaEDef and $\ensuremath{m_{ES} = 
\sqrt{(s/2+\vec{p}_B \cdot \vec{p}_{beams})^2/E_{beams}^2 - \vec{p}_B^{\,2}}}$ 
are used in a fit to provide discrimination between signal and background 
decays. $\sqrt{s}$ is the center-of-mass energy of the colliding particles. 
Here, $P_B = P_{meson}+P_{\ell}+P_{\nu}$ must be evaluated in the laboratory 
frame. We only retain candidates with \FitRegion, thereby removing from the fit
a region with large backgrounds. Fewer than 6.6\% (12.5\%, 7.2\%, 7.4\%, 1.9\%)
of all $\pi^-\ell\nu$ ($\pi^0\ell\nu$, $\omega\ell\nu$, $\eta\ell\nu$, 
$\eta^{\prime}\ell\nu$) events have more than one candidate per event. For 
events with multiple candidates, only the candidate with the largest value of 
$\cos\theta_{\ell}$ is kept. The signal event reconstruction efficiency varies 
between 6.1\% and 8.5\% for \pilnu\ decays, between 2.8\% and 6.0\% for 
\pizlnu\ decays, between 1.0\% and 2.2\% for \omegalnu\ decays, and between
0.9\% and 2.6\% for \etalnu\ decays ($\gamma\gamma$ channel), depending on the 
value of \qq. The efficiency is 0.6\% for both \etalnu\ ($\pi^+\pi^-\pi^0$ 
channel) and \etaplnu\ decays. The efficiencies are given as a function of \qq\
in Tables~\ref{picerror}-\ref{etaerror} of the Appendix.

\section{Backgrounds and Signal Extraction}

 Backgrounds can be broadly grouped into three main categories: decays arising 
from \ulnu\ transitions (other than the signal), decays in \obb\ events 
(excluding \ulnu) and decays in continuum events. The ``\obb'' background is 
the sum of different contributions, where more than $75$\% are from $B 
\rightarrow D/D^*/D^{**}$ decays. For the \pilnu, \pizlnu\ and combined 
\piclnu\ modes, for which there is a large number of candidate events, each of 
the first two categories of background is further split into a background 
category where the pion and the lepton come from the decay of the same $B$ 
meson (``same-$B$'' category), and a background category where the pion and the
lepton come from the decay of different $B$ mesons (``both-$B$'' category). 

\begin{figure*}
\begin{center}
\epsfig{file=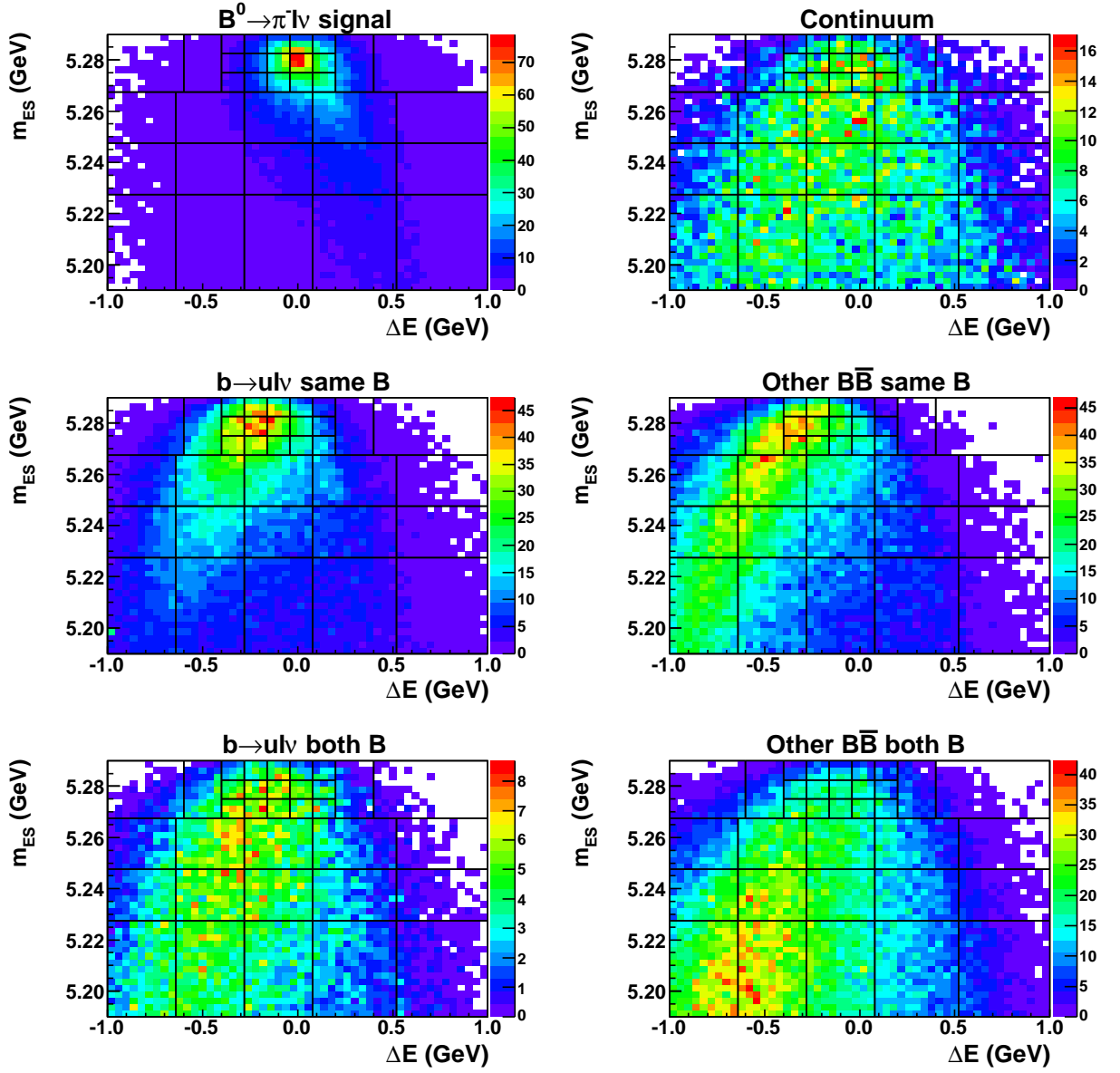,height=16cm}
\caption[]
{\label{dEmES} 
(color online) \DeltaE-\mes\ MC distributions, summed over all bins of \qqr, 
for the six categories of events used in the signal extraction fit, after all 
the selections have been applied, in the case of the \pilnu\ decay channel. 
Also shown is the binning used for this decay mode.}
\end{center}
\end{figure*}

 We use the \DeltaE-\mes\ histograms, obtained from the MC simulation, as 
two-dimensional probability density functions (PDFs) in an extended binned 
maximum-likelihood fit~\cite{Barlow} to the data to extract the yields
of the signal and backgrounds as a function of \qqr. This fit method 
incorporates the statistical uncertainty from the finite MC sample size into 
the fit uncertainty. The \DeltaE-\mes\ plane is subdivided into 34 bins for 
each bin of \qq\ in the fits to the $\pi^-\ell^+\nu$, $\pi^0\ell^+\nu$ and 
$\pi\ell^+\nu$ candidate data where we have a reasonably large number of 
events, and into 19 bins in the fits to the $\omega\ell^+\nu$, $\eta\ell^+\nu$ 
and $\eta^{\prime}\ell\nu$ decay data. The \DeltaE-\mes\ distributions for the 
\pilnu\ decay channel are shown in Fig.~\ref{dEmES}. The binning used in this 
case is also displayed in the figure. We use variable bin sizes because we want
to have a large number of small bin sizes in the signal-enhanced region to 
better define this specific region. The signal-enhanced region is the region of
the \DeltaE-\mes\ plane with a large proportion of signal events. It is 
delimited in our work by the boundaries: $-0.16 < \Delta E < 0.20$ \gev and 
\mes\ $>$ 5.268 \gev\ (see Fig.~\ref{dEmES}). To allow the fit to converge 
quickly we cannot have too many bins in the overall \DeltaE-\mes\ plane. Hence 
the bins outside the signal-enhanced region will have a larger size. The actual
size is dictated by the need to have a good description of the smooth 
backgrounds. The parameters of the fit are the scaling factors of the MC PDFs, 
{\it i.e.}, the factors used to adjust the number of events in a PDF to 
minimize the $\chi^2$ value of the fit.  

 Given the sufficient number of events for the \pilnu\ and combined \piclnu\
decay modes, the data samples can be subdivided in $12$ bins of \qqr\ for the 
signal and two bins for each of the five background categories. The use of two 
bins for each background component allows the fit to adjust for inaccuracies in
the modelling of the shape of the background \qq\ spectra. The boundaries of 
the two background bins of \qqr\ for the \pilnu\ and \piclnu\ decays are: 
[0-18-26.4] \gevsq\ for the \ulnu\ same-$B$ category, [0-22-26.4] \gevsq\ for 
the \ulnu\ both-$B$ category, [0-10-26.4] \gevsq\ for the \obb\ same-$B$ 
category, [0-14-26.4] \gevsq\ for the \obb\ both-$B$ category and [0-22-26.4] 
\gevsq\ for the continuum category. In each case, the \qq\ ranges of the two 
bins are chosen to contain a similar number of events. In the fit to the data, 
we determine for each bin of \qq, the signal yield, the \ulnu, the other 
$B\bar{B}$ and the continuum background yields in each bin of \DeltaE-\mes.
 
Note, however, that the scaling factors obtained for each background are 
constrained to have the same value over their ranges of \qq\ defined above. We 
thus have a total of 22 parameters and $(12 \times 34 - 22)$ degrees of freedom
in the fit to the \pilnu\ data and to the combined \piclnu\ data. The limited 
number of events for the other signal modes reduces the number of parameters, 
and hence the number of \qq\ bins, that can be used for  the fits to converge. 
Table~\ref{BinningMode} shows the number of bins of \qq\ used for each signal 
mode as a function of the fit category.

\begin{table}[t]
\caption[]{\label{BinningMode} Categories and number of fit parameters for each
decay mode.}
\begin{center}
\begin{tabular}{cccccc}
\hline\hline
Categories &                  \multicolumn{5}{c}{Decay mode}   \\
  & $\pi^- \ell \nu$ & $\pi^0 \ell \nu$ & $\omega \ell \nu$ & $\eta \ell \nu$ 
($\gamma\gamma$) & $\eta^{\prime} \ell \nu$ ($\gamma\gamma$)\\
  & $\pi \ell \nu$ & & & $\eta \ell \nu$ ($\gamma\gamma$ \& $3\pi$) & $\eta 
\ell \nu$ ($3\pi$) \\ 
\hline
Signal          & 12 & 11 & 5 & 5 & 1  \\ \hline
\ulnu\ same $B$ & 2  &  1 & \multirow{2}{*}{1} & \multirow{2}{*}{fixed} & 
\multirow{2}{*}{fixed}  \\ 
\ulnu\ both $B$ & 2  &  1 &  &  &  \\ \hline
\obb\ same $B$  & 2  &  1 & \multirow{2}{*}{1} & \multirow{2}{*}{1} & 
\multirow{2}{*}{1} \\ 
\obb\ both $B$  & 2  &  1 &  &  &  \\ \hline
Continuum       & 2  &  1 & 1 & 1 & fixed \\ \hline\hline
\end{tabular}
\end{center}
\end{table}

\begin{figure*}
\begin{center}
\epsfig{file=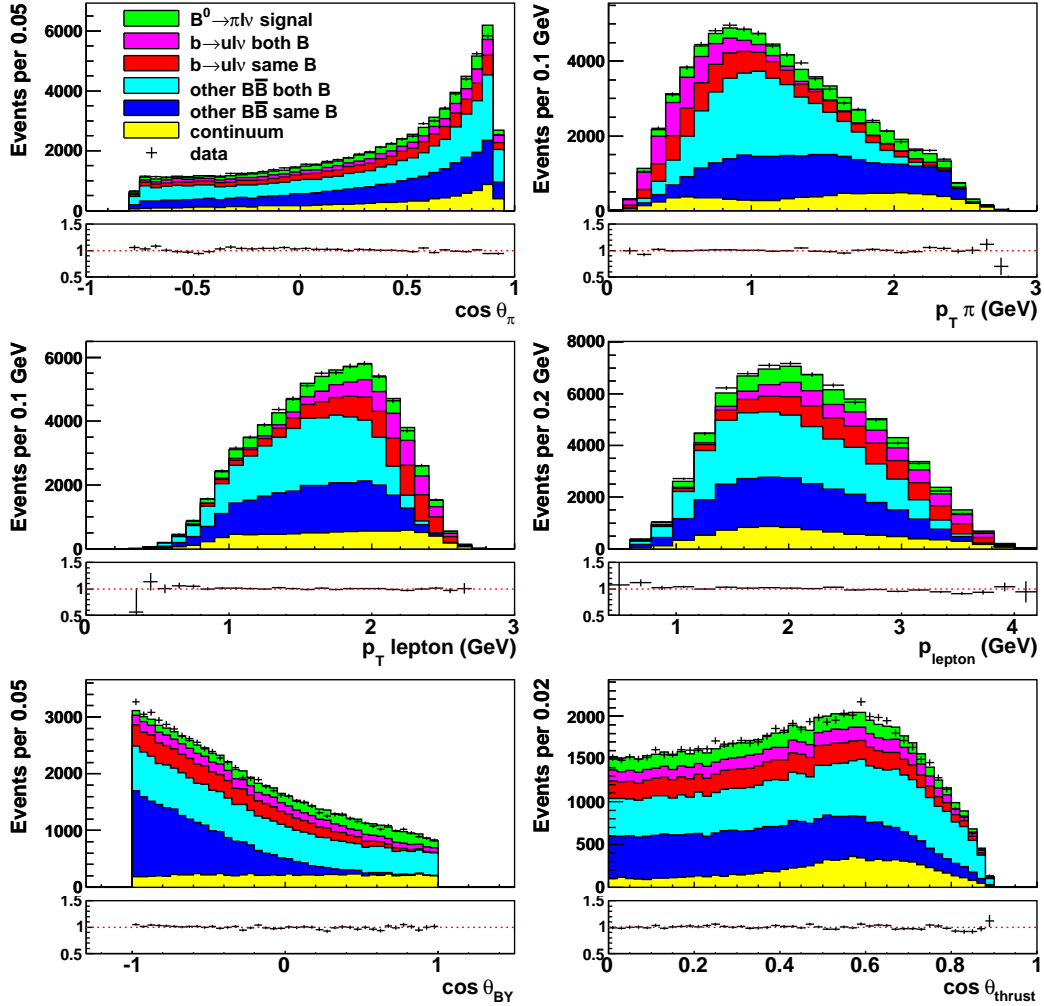,height=13.5cm}
\caption[]{\label{DatMC1Pi} (color online) Comparison of the on-resonance data 
and MC simulation, for \pilnu\ decays, after all analysis cuts and MC 
simulation corrections have been applied. The $Y$ signal candidates related 
distributions are  generated from events in the \DeltaE\ and \mes\ plane with 
the signal-enhanced region removed. The ratios of data/MC events are presented 
below each panel. The general level of agreement is better than 5\%.}
\end{center}
\end{figure*}

\begin{table*}
\caption[]{\label{yieldBGtable} Fitted yields in the full \qq\ range 
investigated for the signal and each background category, total fitted yield
and experimental data events, and values of $\chi^2$ for the overall fit 
region.} 
\begin{center}
\begin{tabular}{lp{0.1cm}cp{0.1cm}cp{0.1cm}cp{0.1cm}cp{0.1cm}cp{0.1cm}c}
\hline\hline
Decay mode          & &$\pi^-\ell^+\nu$& &$\pi^0\ell^+\nu$& &$\pi\ell^+\nu$& &$\omega\ell^+\nu$& &$\eta\ell^+\nu$& &$\eta^{\prime}\ell^+\nu$ \\ \hline
Signal              & & $9297\pm 316$& &$3204\pm 170$& &$12448\pm 361$& &$1861\pm 233$& &$867\pm 101$    & &$141\pm 49$ \\
\ulnu\              & & $15689\pm 664$& &$7810\pm 334$& &$23284\pm 796$& &$3246\pm 293$& &$2411 (fixed)$& &$242 (fixed)$ \\   
Other $B\bar{B}$& & $44248\pm 656$& &$10795\pm 307$& &$55350\pm 777$& &$8778\pm 246$& &$11167\pm187$ & &$2984\pm87$ \\ 
Continuum           & & $9159\pm 459$ & &$4173\pm 236$& &$13283\pm 537$& &$2776\pm 270$& &$2505\pm155$  & &$493 (fixed)$ \\
\hline
Fitted yield  & & $78393\pm 507$& &$25982\pm 228$& &$104365\pm 531$& &$16661\pm 172$& &$16950\pm 153$ & &$3860\pm 71$ \\
Data events         & & $78387\pm280$& &$25977\pm 161$& &$104364\pm323$& &$16662\pm 129$& &$16901\pm 130$ & &$3857\pm62$  \\
$\chi^2$/ndf        & & $385.3/386$& &$324.9/358$& &$387.7/386$& &$74.9/87$& &$100.1/88$& &$16.8/17$ \\
\hline\hline
\end{tabular}
\end{center}
\end{table*}

  As an initial estimate in the fit, the MC continuum background yield and 
\qq-dependent shape are first normalized to match the yield and \qq-dependent 
shape of the off-resonance data control sample. This results in a large 
statistical uncertainty due to the small number of events in the off-resonance
data. To improve the statistical precision, the continuum background is allowed
to vary in the fit to the data for the $\pi\ell\nu$, $\omega\ell\nu$ and 
$\eta\ell\nu (\gamma\gamma)$ modes where we have a relatively large number of 
events. The fit result is compatible with the measured distribution of 
off-resonance data. Whenever a background is not varied in the fit, it is fixed
to the MC prediction, except for the continuum background which is fixed to its
normalized yield and \qq-dependent shape using the off-resonance data. The 
background parameters which are free in the fit, typically require an 
adjustment of less than 10\% with respect to the MC predictions. The initial 
agreement between MC and data is already good before we do any fit. After the 
fit, the agreement becomes excellent, as can be seen in Fig.~\ref{DatMC1Pi} for
a number of variables of interest. The values of the scaling factors, obtained 
in this work, are presented in Table~\ref{ScalingF} of the Appendix for each 
decay channel. The full correlation matrices of the fitted scaling factors are 
given in Tables~\ref{table1}-\ref{table8} of the Appendix.
 
 We refit the data on several different subsets obtained by dividing the final 
data set based on time period, electron or muon candidates, by modifying the 
\qq, \DeltaE\ or \mes\ binnings, and by varying the event selections. We obtain
consistent results for all subsets.  We have also used MC simulation to verify 
that the nonresonant decay contributions to the resonance yields are 
negligible. For example, we find that there are 30 nonresonant $\pi^+\pi^-\pi^0
\ell\nu$ events out of a total yield of $1861\pm233$ events for the \omegalnu\ 
decay channel.

\begin{figure}
\begin{center}
\epsfig{file=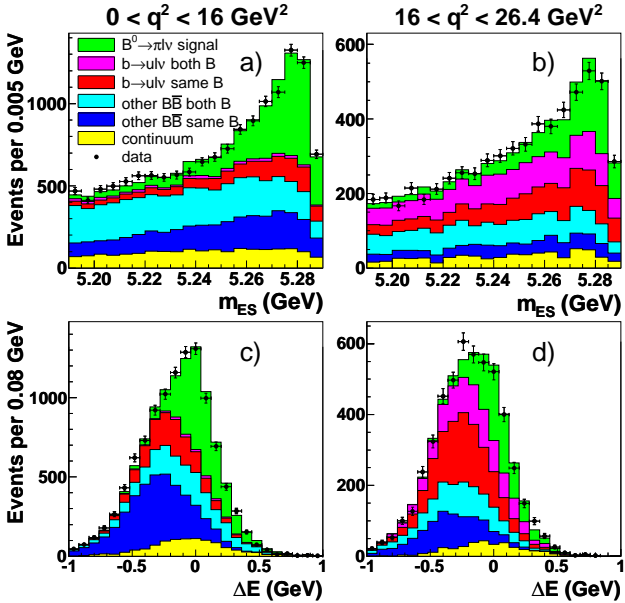,height=8.0cm}
\caption[]{\label{dEmESProjPilnu} (color online) Projections of the data and 
fit results for the \pilnu\ decays, in the 
signal-enhanced region: (a,b) \mes\ with $-0.16 < \Delta E < 0.20$ \gev; and 
(c,d) \DeltaE\ with \mes\ $>$ 5.268 \gev. The distributions (a,c) and (b,d) are
projections for \qqr\ $<$ 16 \gevsq\  and for \qqr\ $>$ 16 \gevsq, 
respectively.}
\end{center}
\end{figure}

\begin{figure}
\begin{center}
\epsfig{file=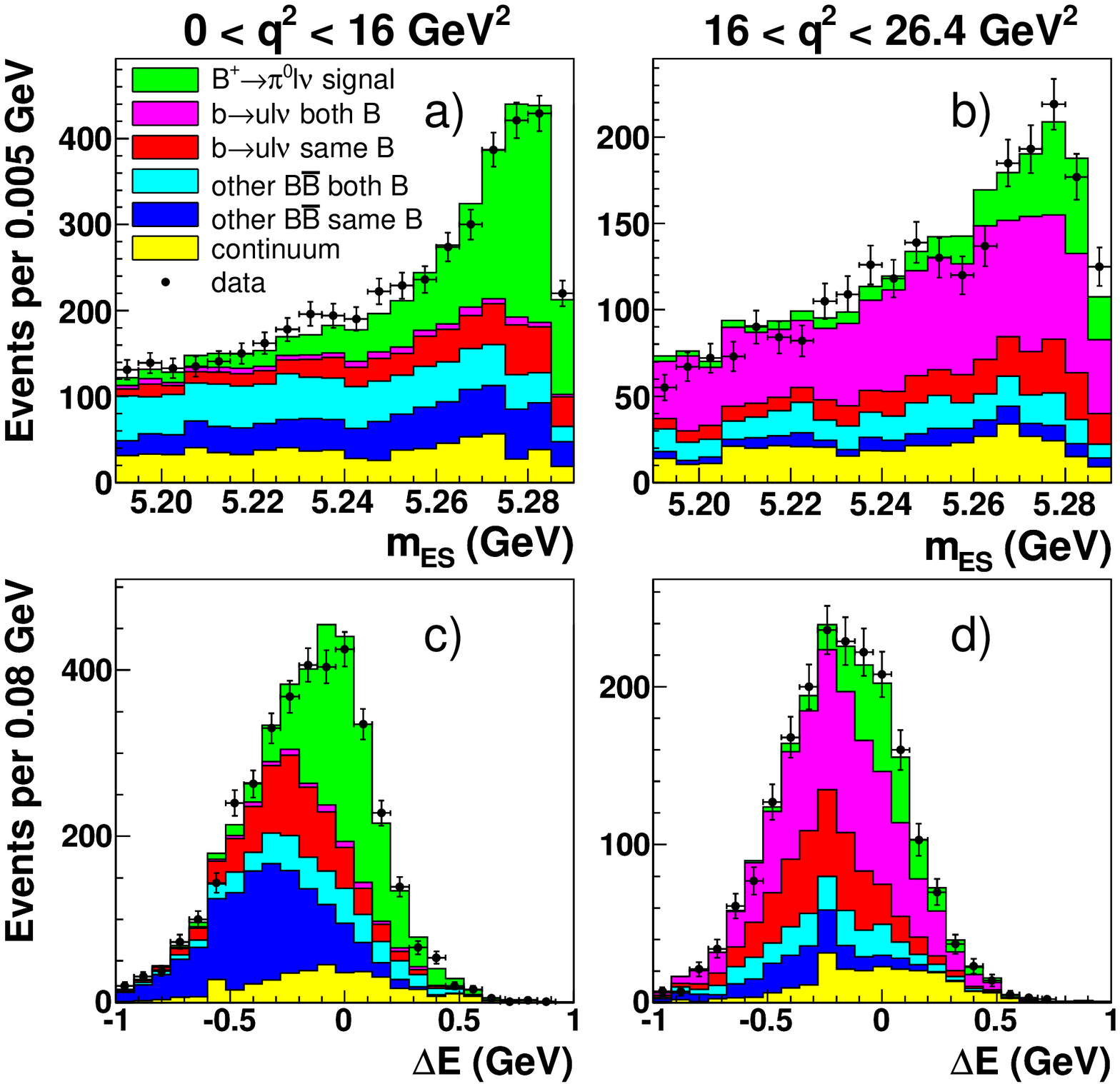,height=8.0cm}
\caption[]{\label{dEmESProjPi0lnu} (color online) Projections of the data and 
fit results for the \pizlnu\ decays, in the 
signal-enhanced region: (a,b) \mes\ with $-0.16 < \Delta E < 0.20$ \gev; and 
(c,d) \DeltaE\ with \mes\ $>$ 5.268 \gev. The distributions (a,c) and (b,d) are
projections for \qqr\ $<$ 16 \gevsq\  and for \qqr\ $>$ 16 \gevsq, 
respectively.}
\end{center}
\end{figure}

\begin{figure}
\begin{center}
\epsfig{file=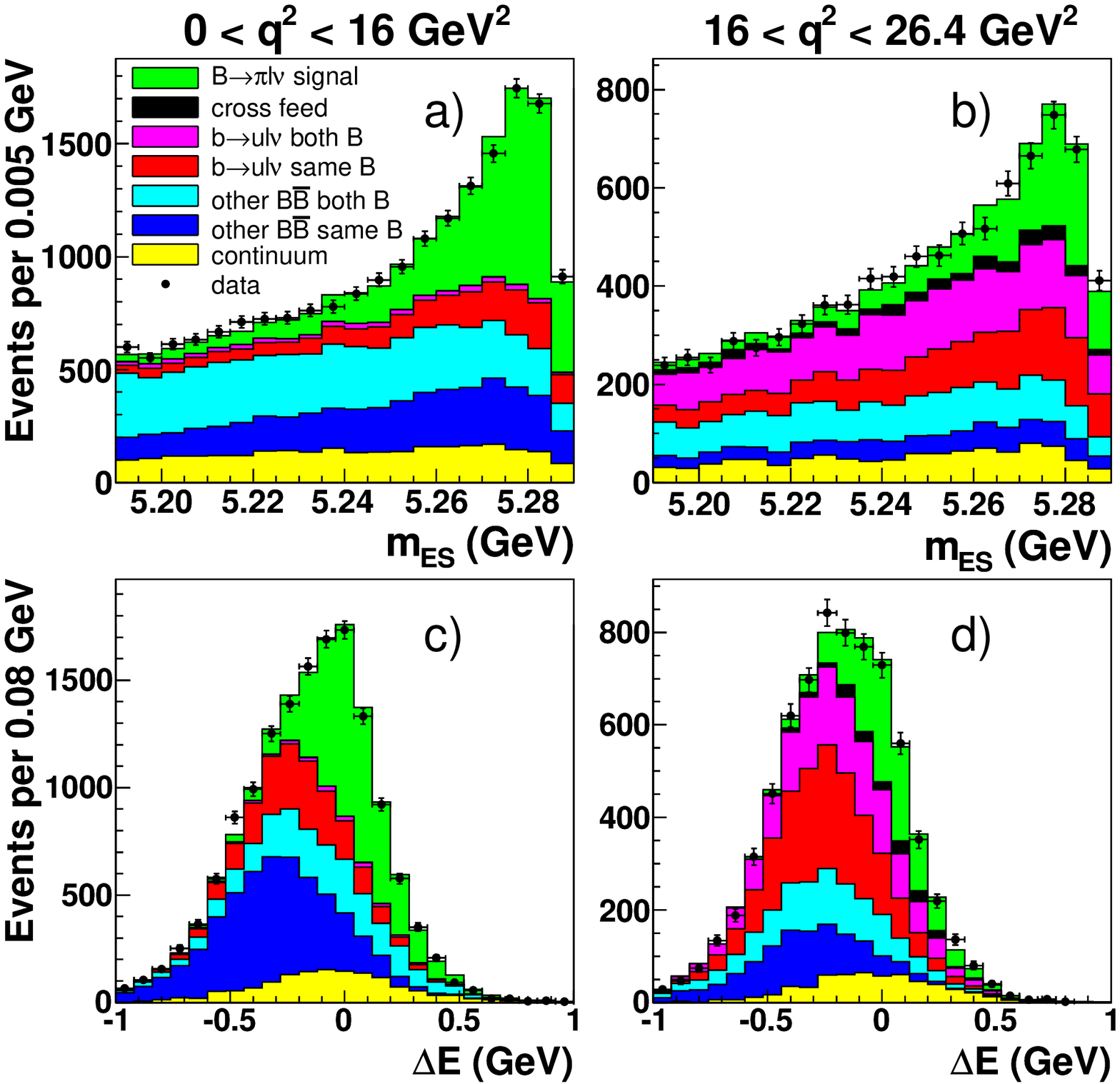,height=8.0cm}
\caption[]{\label{dEmESProjPiclnu} (color online) Projections of the data and 
fit results for the combined \pilnu\ and \pizlnu\ decays, in the 
signal-enhanced region: (a,b) \mes\ with $-0.16 < \Delta E < 0.20$ \gev; and 
(c,d) \DeltaE\ with \mes\ $>$ 5.268 \gev. The distributions (a,c) and (b,d) are
projections for \qqr\ $<$ 16 \gevsq\  and for \qqr\ $>$ 16 \gevsq, 
respectively.}
\end{center}
\end{figure}

\begin{figure*}
\begin{center}
\epsfig{file=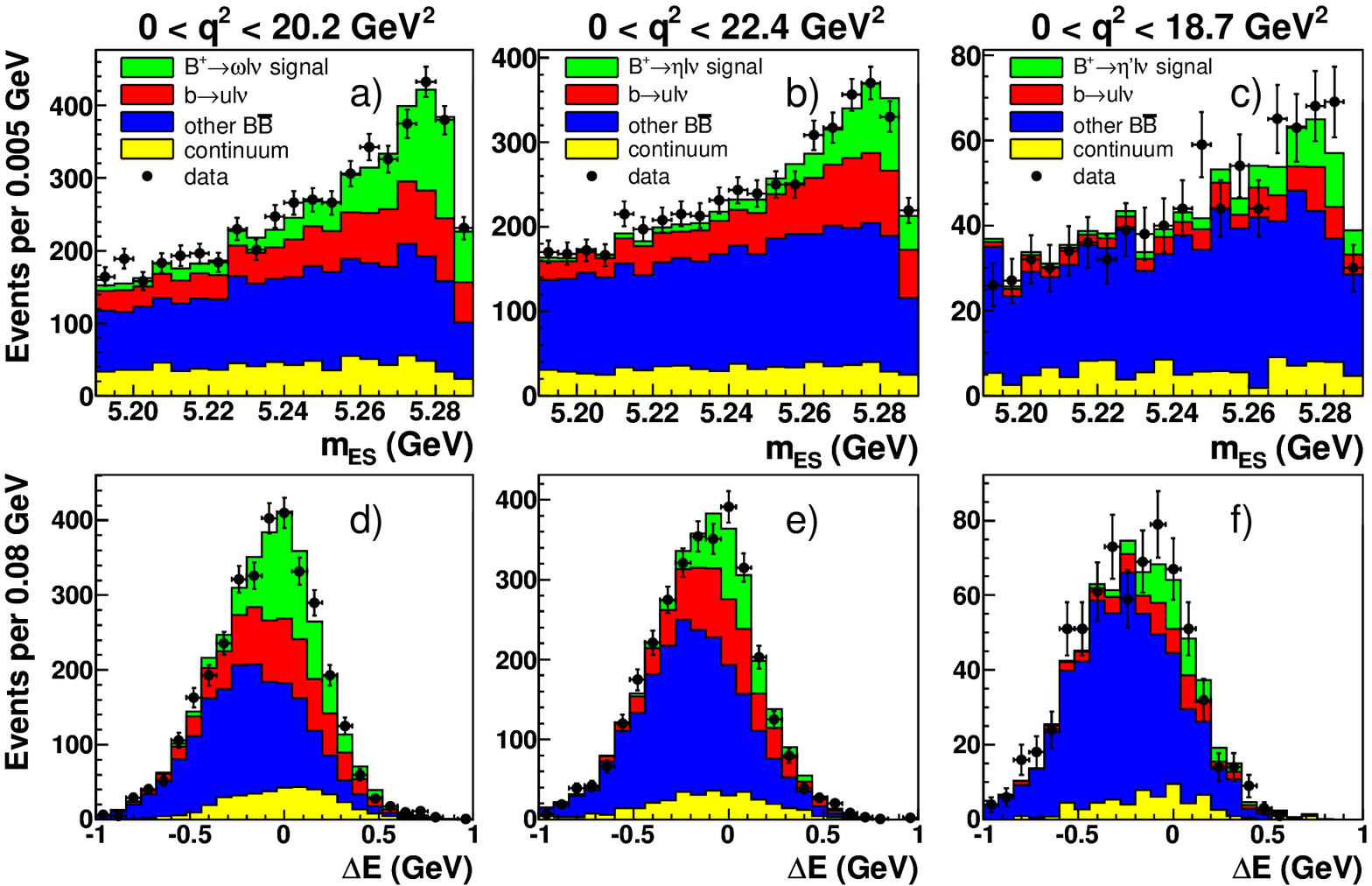,height=8.0cm}
\caption[]{\label{dEmESProjEta} (color online) Projections of the data and fit 
results for the \omegalnu\ and \etaetaplnu\ decays, in the signal-enhanced 
region: (a,b,c) \mes\ with $-0.16 < \Delta E < 0.20$ \gev; and (d,e,f) \DeltaE\
with \mes\ $>$ 5.268 \gev. The distributions (a,d), (b,e) and (c,f) are 
projections for the \omegalnu, combined \etalnu, and \etaplnu\ decays, 
respectively.}
\end{center}
\end{figure*}

 For illustrative purposes only, we show in Figs.~\ref{dEmESProjPilnu}, 
~\ref{dEmESProjPi0lnu}, and~\ref{dEmESProjPiclnu}, \DeltaE\ and \mes\ fit 
projections in the signal-enhanced region for the \pilnu, \pizlnu\ and combined
\pilnu\ and \pizlnu\ decays, respectively, in two ranges of \qq\ corresponding 
to the sum of eight bins below and four bins above \qqr\ = 16 \gevsq, 
respectively. More detailed \DeltaE\ and \mes\ fit projections in each \qqr\ 
bin are shown in Figs.~\ref{dEFitProjDataPic} and~\ref{mESFitProjDataPic} of 
the Appendix for the combined \piclnu\ decays. The data and the fit results are
in good agreement. Fit projections for \omegalnu\ and \etaetaplnu\ decays, over
their \qq\ ranges of investigation, are shown in Fig.~\ref{dEmESProjEta}. Table
\ref{yieldBGtable} gives the fitted yields in the full \qq\ range studied
for the signal and each background category as well as the $\chi^2$ values and 
degrees of freedom for the overall fit region. The yield values in the \etalnu\
column are the result of the fit to the combined $\gamma\gamma$ and $3\pi$ 
modes. 

\section{Systematic Uncertainties}

 Systematic uncertainties on the values of the partial branching fractions, 
\bfpilnuqq, and their correlations among the \qq\ bins have been investigated. 
These uncertainties are estimated from the variations of the resulting partial 
BF values (or total BF values for \etaplnu\ decays) when the data are 
reanalyzed by reweighting different simulation parameters such as BFs and form 
factors. For each parameter, we use the full MC dataset to produce new 
\DeltaE-\mes\ distributions (``MC event samples'') by reweighting the parameter
randomly over a complete Gaussian distribution whose standard deviation is 
given by the uncertainty on the parameter under study. One hundred such samples
are produced for each parameter. Each MC event sample is analyzed the same way 
as real data to determine values of \bfpilnuqq\ (or total BF values for 
\etaplnu\ decays). The contribution of the parameter to the systematic 
uncertainty is given by the RMS value of the distribution of these \bfpilnuqq\ 
values over the one hundred samples.

\begin{table*}
\caption[]{\label{errorsPi} Values of signal yields, \bfpilnuqq\ and their 
relative uncertainties (\%) for \pilnu\ and \pizlnu\ decays.}
\begin{center}
\begin{tabular}{lp{0.1cm}cp{0.1cm}cp{0.1cm}cp{0.1cm}cp{1cm}cp{0.1cm}cp{0.1cm}cp{0.1cm}c} 
\hline\hline Decay mode & & \multicolumn{7}{c}{$\pi^-\ell^+\nu$} & & \multicolumn{7}{c}{$\pi^0\ell^+\nu$}  \\ 
\hline \qq\ range (\gevsq) & &\qq$<$12  & &\qq$<$16 & &\qq$>$16 & & $0<q^2<26.4$ & &\qq$<$12  & &\qq$<$16 & &\qq$>$16 & & $0<q^2<26.4$ \\ 
\hline
Unfolded yield 	             &    & 5604.1 & &6982.4& & 2314.2& & 9296.5 & & 2231.7 & &2666.7& & 537.3& & 3204.1 \\ 
\bfpilnuqq\ ($10^{-4}$)               &    & 0.83   & & 1.07 & & 0.40  & & 1.47    & & 0.46   & & 0.61 & & 0.16  & & 0.77 \\
\hline
Statistical error            &    & 4.3    & & 3.8 & & 6.7 & & 3.5        & & 6.6    & &  5.3 & & 17.8 & & 5.7 \\ 
Detector effects             &    & 3.4    & & 3.5 & & 3.2 & & 2.8        & & 2.9    & &  2.8 & & 3.0 & & 2.6 \\ 
Continuum bkg                &    & 0.4    & & 0.4 & & 1.4 & & 0.4        & & 1.2    & &  0.8 & & 7.1 & & 1.1 \\ 
\ulnu\ bkg& & 1.6    & & 1.4 & & 2.1 & & 1.3        & & 1.7    & &  1.5 & & 5.9 & & 1.9 \\ 
\clnu\ bkg& & 0.6    & & 0.5 & & 0.6 & & 0.5        & & 0.6    & &  0.4 & & 1.0 & & 0.4 \\ 
Other effects                &    & 2.2    & & 2.1 & & 2.1 & & 2.1        & & 2.1    & &  2.1 & & 2.5 & & 2.0 \\ 
\hline
Total uncertainty            &    & 6.2    & & 5.8 & & 8.1 & & 5.1       & & 7.9   & &  6.5 & & 20.4 & & 6.9 \\ 
\hline\hline 
\end{tabular}
\end{center}
\end{table*}

\begin{table*}
\caption[]{\label{errorsAll} Values of signal yields, \bfpilnuqq\ and their 
relative uncertainties (\%) for combined \piclnu, \omegalnu, combined \etalnu\
($\gamma\gamma$ and $3\pi$ decay channels) and \etaplnu\ decays.}
\begin{center}
\begin{tabular}{lp{0.1cm}cp{0.1cm}cp{0.1cm}cp{0.1cm}cp{0.5cm}cp{0.5cm}cp{0.5cm}c} 
\hline\hline Decay mode & & \multicolumn{7}{c}{combined $\pi\ell^+\nu$} & & $\omega\ell^+\nu$ & & $\eta\ell^+\nu$ & & $\eta^{\prime}\ell^+\nu$ \\ 
\hline \qq\ range (\gevsq) & &\qq$<$12  & &\qq$<$16 & &\qq$>$16 & & $0<q^2<26.4$ & & $0<q^2<20.2$ & & $0<q^2<22.4$ & & $0<q^2<18.7$ \\ 
\hline
Unfolded yield 		     &    & 7805.4 & &9618.9& & 2829.0& & 12447.9 & & 1860.8 & & 867.3 & & 141.1\\
\bfpilnuqq\ ($10^{-4}$)               &    & 0.83  & & 1.08 & & 0.37  & & 1.45     & &  1.19 & &  0.38 & & 0.24 \\
\hline
Statistical error            &    & 3.6  & & 3.2 & & 5.8 & & 3.0          & &  13.0 & &  13.7 & & 34.9 \\ 
Detector effects             &    & 3.7  & & 3.8 & & 3.5 & & 3.1          & &   3.9 & &   9.8 & & 7.7 \\ 
Continuum bkg                &    & 0.4  & & 0.6 & & 3.3 & & 0.6          & &   3.2 & &   -   & & 5.8  \\ 
\ulnu\ bkg& & 1.6  & & 1.4 & & 4.0 & & 1.4          & &   5.1 & &   8.4 & & 4.9  \\ 
\clnu\ bkg& & 0.4  & & 0.4 & & 0.4 & & 0.3          & &   1.0 & &   2.1 & & 3.3  \\ 
Other effects                &    & 1.8  & & 1.7 & & 1.5 & & 1.6          & &   1.8 & &   1.8 & & 2.4  \\ 
\hline
Total uncertainty            &    & 5.8  & & 5.5 & & 8.7 & & 4.9         & &  15.0 & &  19.0 & & 36.7 \\ 
\hline\hline 
\end{tabular}
\end{center}
\end{table*}

  The systematic uncertainties due to the imperfect description of the detector
in the simulation are computed by using the uncertainties determined from 
control samples. These include the tracking efficiency of all charged particle 
tracks, the particle identification efficiencies of signal candidate tracks,
the calorimeter efficiencies (varied separately for photons and \KL), the 
energy deposited in the calorimeter by \KL mesons as well as their production 
spectrum. The reconstruction of these neutral particles affects the analysis 
through the neutrino reconstruction used to obtain the values of $\DeltaE$
and $m_{ES}$. 

 The uncertainties due to the generator-level inputs to the simulation are 
given by the uncertainties in the BFs of the background \ulnu\ and \clnu\ 
processes, in the BFs of the secondary decays producing leptons, and in the BFs
of the $\Upsilon(4S) \rightarrow B\bar{B}$ decays~\cite{PDG10}. The $B 
\rightarrow X \ell\nu$ form-factor uncertainties, where $X = (\pi,\rho,\omega,
\eta^{(\prime)},D,D^*, D^{**})$, are given by recent calculations or 
measurements~\cite{PDG10}. The uncertainties in the heavy quark parameters used
in the simulation of nonresonant \ulnu\ events are given in 
Ref.~\cite{Henning}. The uncertainty due to final state radiation (FSR) 
corrections calculated by PHOTOS~\cite{photos} is given by 
20\%~\cite{photosErr} of the difference in the values of the BF obtained with 
PHOTOS switched on and with PHOTOS switched off. The uncertainty due to the 
modelling of the continuum is obtained by comparing the shape of its \qqr\ 
distribution to that of the off-resonance data control sample. When the 
continuum is fixed in the fit, the uncertainty in the total yield is used 
instead. The uncertainty in that case is given by the comparison of the MC 
total yield to the one measured off-resonance. Finally, the uncertainty due to 
$B$ counting has been established to be 0.6\% in \babar.

  Additional details on the various sources of systematic uncertainties 
considered in this analysis are presented in Ref.~\cite{Jochen}. The individual
sources are, to a good approximation, uncorrelated. Their associated 
contributions to the uncertainties can therefore be added in quadrature to 
yield the total systematic uncertainties for each decay mode. 

  The list of all the systematic uncertainties, as well as their values for the
partial and total BFs, are given in Tables~\ref{picerror}-\ref{etaerror} of the
Appendix. The term ``Signal MC stat error'' in these tables incorporates the 
systematic uncertainty due to the unfolding procedure. The correlation matrices
obtained in the measurement of the partial BFs are presented in 
Tables~\ref{StatCovPic}-\ref{CovEta}. Condensed versions of all the 
uncertainties, together with signal yields and partial BFs in selected \qqr\ 
ranges, are given in Table~\ref{errorsPi} for the \pilnu\ and \pizlnu\ decays, 
and in Table~\ref{errorsAll} for the combined \piclnu\ decays, as well as for 
the \omegalnu\ and \etaetaplnu\ decays. The values given for the \etalnu\ 
decays are those obtained from the combined fit to the distributions of the 
$\eta \rightarrow \gamma \gamma$ and $\eta \rightarrow \pi^+ \pi^- \pi^0$ 
channels. The ranges of \qqr\ delimited by the numbers $12, 16$ are ranges used
in theoretical predictions. We also give the results for the fully allowed 
kinematical range of \qqr.

\section{Branching Fraction Results}

 \begin{table}
\caption[]{\label{BF} Values of the total branching fractions obtained in this
analysis and previous results. The two uncertainties are statistical and 
systematic, respectively. All BF values are $\times 10^{-4}$.}
\begin{center}
\begin{tabular}{lp{0.05cm}cp{0.05cm}cp{0.05cm}c}
\hline\hline
Decay mode && This analysis && Previous results && Ref. \\ \hline
\piclnu\ && $1.45\pm 0.04 \pm 0.06$ && $1.41\pm 0.05 \pm 0.07$ && \cite{Jochen} \\
\pilnu\  && $1.47\pm 0.05 \pm 0.06$ && $1.44\pm 0.06 \pm 0.07$ && \cite{Jochen} \\
         &&                         && $1.42\pm 0.05 \pm 0.07$ && \cite{Simard} \\
         &&                         && $1.49\pm 0.04 \pm 0.07$ && \cite{Belle} \\
\pizlnu\ && $0.77\pm 0.04 \pm 0.03$ && $0.76\pm 0.06 \pm 0.06$ && \cite{Jochen} \\
\omegalnu\ && $1.19\pm 0.16\pm 0.09$ && $1.21\pm 0.14\pm 0.10$ && \cite{Wulsin} \\
\etalnu\ && $0.38\pm 0.05 \pm 0.05$ && $0.36\pm 0.05 \pm 0.04$ && \cite{Simard} \\
\etaplnu\ && $0.24\pm 0.08 \pm 0.03$ && $0.24\pm 0.08\pm 0.03$ && \cite{Simard} \\
\hline\hline
\end{tabular}
\end{center}
\end{table}

 The total BF for the \etaplnu\ decays and the partial BFs for the other four 
decay modes are calculated using the unfolded signal yields, the signal 
efficiencies given by the simulation and the branching fractions \bfupsbzbz\ 
$= 0.484\pm0.006$ and \bfupsbpbm\ $= 0.516\pm0.006$~\cite{PDG10}. The values of
the total BF obtained in this work are compared in Table~\ref{BF} to those 
reported recently. 

 The BFs for the \etaplnu\ and \etalnu\ decays are consistent with those 
presented in our earlier work~\cite{Simard} even though there are significant 
differences between the two analyses. We now use updated BFs and form-factor 
shapes; we have tightened various selections; we have subdivided the data in 
five signal bins for the \etalnu\ decays compared to the previous three bins; 
and we have also investigated the full kinematically allowed ranges of \qqr\ 
whereas this range was earlier restricted to less than 16 \gevsq\ due to the 
very large backgrounds at high $q^2$. Thus, the present BF values supersede the
earlier ones~\cite{Simard}. It should be noted that the total BF value for the 
\etaplnu\ decays has a significance of $3.2\sigma$ when we take into account 
only the statistical uncertainty~\cite{signif}. Taking into account the effect 
of the systematic uncertainty which increases the total uncertainty by about 
3\% leads to a reduced significance of $3.1\sigma$. We find that the total BF 
of the \etalnu, $\eta \rightarrow \gamma\gamma$ decays 
(($0.36\pm0.06\pm0.05)\times 10^{-4}$) is compatible with the total BF measured
for the \etalnu, $\eta \rightarrow \pi^+\pi^-\pi^0$ decays 
(($0.46\pm0.10\pm0.05)\times 10^{-4}$). The total BF value in Table~\ref{BF} 
for \etalnu\ decays is obtained from a fit to the combined $\gamma\gamma$ and 
$3\pi$ decay channels. This value is in good agreement with the weighted 
average of the total BFs obtained separately for these two decay channels. 

 The present BF value for \omegalnu\ decays is in good agreement with our
previous result~\cite{Wulsin}, as shown in Table ~\ref{BF}. In the present 
analysis, we have a larger number of $\omega\ell\nu$ events ($1861\pm 233$ 
compared to $1125\pm131$ in Ref.~\cite{Wulsin}) and a better signal/background 
ratio (12.6\% versus 9.4\%). We now have a slightly larger statistical 
uncertainty because some of the backgrounds were previously fixed while we now 
fit them to the data. On the other hand, this different treatment of the 
backgrounds leads to a smaller systematic uncertainty in the present case. 
Another difference arises in the treatment of the combinatoric background, 
which is subtracted in Ref.~\cite{Wulsin} using a fit to the mass sideband 
data, while it is part of the likelihood fit in the present study. The other 
important difference is the use of \qq\ bins of equal width in this analysis 
compared with varied bin width in Ref.~\cite{Wulsin}. In addition, our yields 
are unfolded to correct for the reconstruction effects on the measured values 
of $q^2$. The results obtained in this work use the same dataset as those of 
Ref.~\cite{Wulsin} but use a different analysis strategy and selection as 
indicated. This results in a small (estimated to be 14\%) statistical overlap 
between the samples and a different sensitivity to sources of systematic 
uncertainty (estimated correlation of 75\%). Since the choice of \qq\ binning 
differs between the two analyses, only the total branching fractions can be 
combined. Accounting for the major sources of correlation between the 
measurements, the combined \bfomegalnu\ result is: 
$(1.20\pm0.11\pm0.09)\times 10^{-4}$.
     
 Table~\ref{BF} lists the fitted branching fractions for \pilnu, \pizlnu\ and 
the combined \piclnu\ modes. The \pizlnu\ result is used to confirm the \pilnu\
result, using the isospin symmetry relation:
\begin{eqnarray}
\text{\bfpilnu} &=& \text{\bfpiZlnu} \times 2\frac{\tau_0}{\tau_+} \nonumber \\
                &=& (1.43\pm 0.08\pm 0.06)\times 10^{-4} \nonumber
\end{eqnarray} 
where $\tau_+/\tau_0 = 1.079 \pm 0.007$~\cite{PDG10} is the ratio of the 
lifetimes of $B^+$ and $B^0$ decays. The value of the branching fraction thus 
obtained is compatible with the BF value obtained directly for the \pilnu\ 
decays (see Table~\ref{BF}). The combined \piclnu\ decays result is based on 
the use of all $\pi\ell\nu$ decay events where the neutral pion events in a 
given \qq\ bin are converted into equivalent charged pion events assuming the 
above isospin symmetry relation to hold for the total yield in each \qq\ bin.
 Using these combined events leads to a smaller statistical uncertainty on the 
BF value. 

 The values of the present total BFs for the combined \piclnu\ decays, the 
\pilnu\ decays and the \pizlnu\ decays are seen to be in good agreement with 
those reported earlier by \babar,~\cite{Jochen,Simard} and Belle~\cite{Belle}. 
However, the present values are based on updated values of BFs and form-factor 
shapes, and a larger data set compared to the earlier 
works~\cite{Jochen,Simard}. In particular, we now have an improved model for 
the hybrid MC~\cite{Henning} distributions that describe the combination of 
resonant and nonresonant \ulnu\ decays. This model entails the use of the BGL 
parametrization for the \piclnu\ decays~\cite{Simard}, the Ball parametrization
for the \omegalnu\ decays and the BK parametrization for the \etaetaplnu\ 
decays, rather than the much older ISGW2~\cite{Isgur} parametrization. The use 
of this model leads to an increase of $3.5\%$ in the total BF value for the 
\pilnu\ decays, going from a value of $1.42 \times 10^{-4}$ as established 
earlier~\cite{Simard} to the present value of $1.47 \times 10^{-4}$. This 
increase of 3.5\% is significant in view of the total uncertainty of 5.1\% 
obtained in the measurement of the total BF. Thus, the present values of BF 
for \piclnu, \pilnu\ and \pizlnu\ decays supersede the earlier 
results~\cite{Jochen,Simard}.

\begin{figure}
\begin{center}
\epsfig{file=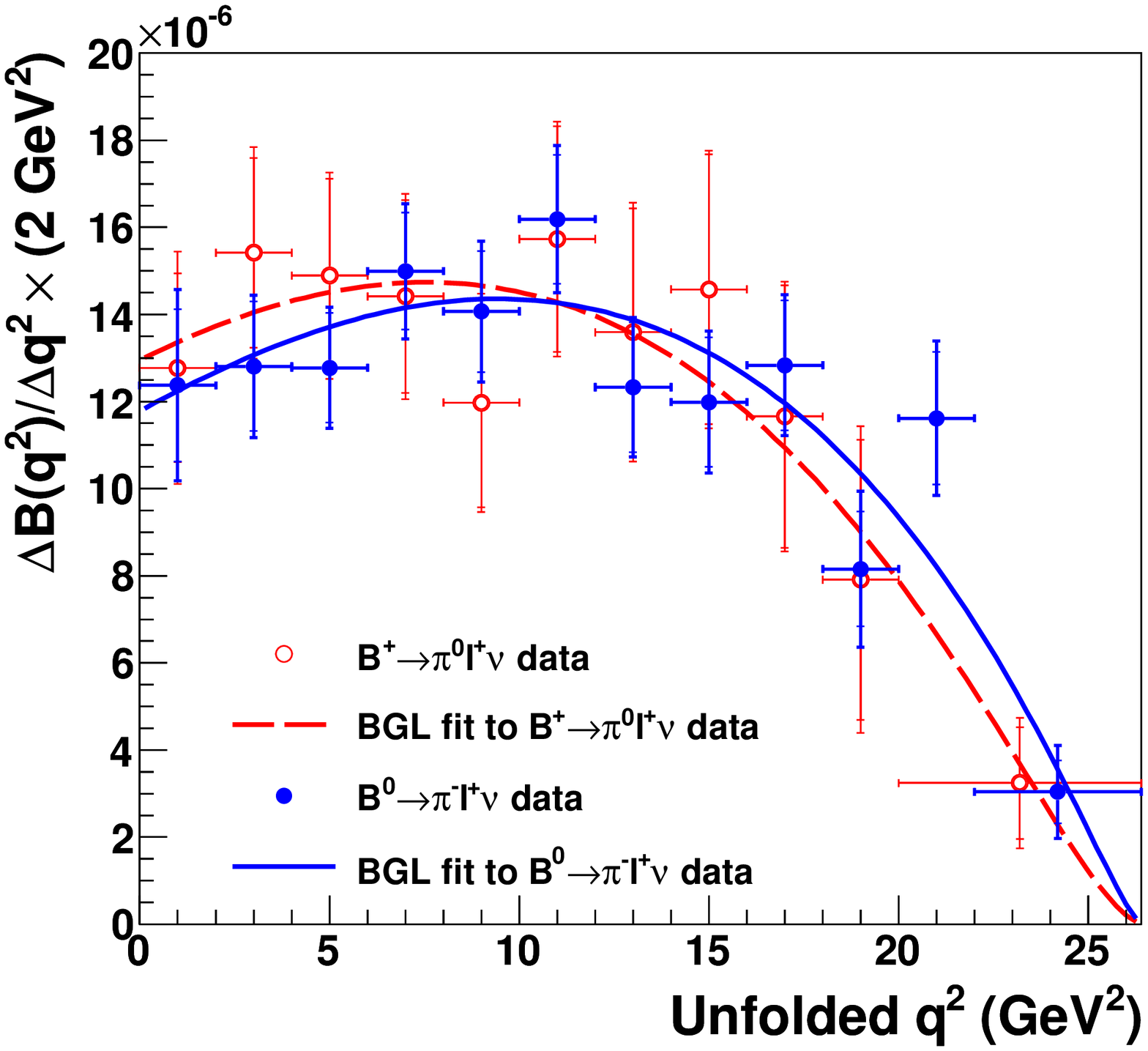,height=8.0cm}
\caption[]{\label{figFplusPi+Pi0} 
(color online) Partial \bfpilnuqq\ spectra in 12 bins of \qq\ for \pilnu\ 
and 11 bins of \qq\ for \pizlnu\ decays. The data points are placed in the middle of each bin whose
width is defined in Table~\ref{pierror}.The smaller error bars are statistical 
only while the larger ones also include systematic uncertainties. The solid 
blue curve shows the result of the fit to the \pilnu\ data of the 
BGL~\cite{BGL} parametrization while the dashed red curve shows the result of 
the fit to the \pizlnu\ data of the same parametrization.}
\end{center}
\end{figure}

\begin{figure}
\begin{center}
\epsfig{file=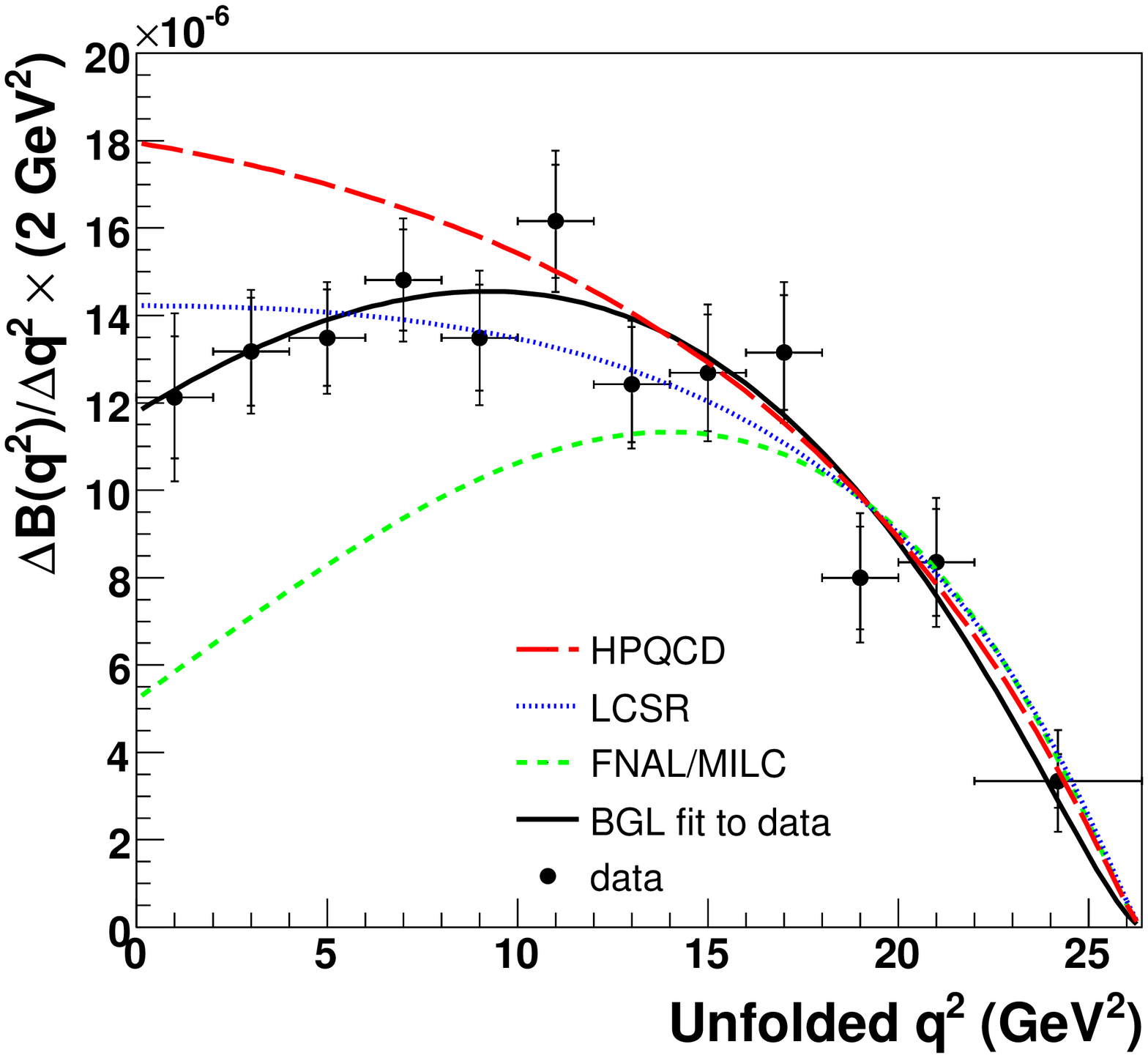,height=8.0cm}
\caption[]{\label{figFplus} 
(color online) Partial \bfpilnuqq\ spectrum in 12 bins of \qq\ for \piclnu\ 
decays. The data points are placed in the middle of each bin whose width is 
defined in Table~\ref{picerror}.The smaller error bars are statistical only 
while the larger ones also include systematic uncertainties. The solid black 
curve shows the result of the fit to the data of the BGL~\cite{BGL} 
parametrization. The data are also compared to unquenched LQCD calculations 
(HPQCD~\cite{HPQCD06}, FNAL~\cite{FNAL}) and a LCSR calculation~\cite{LCSR2}.}
\end{center}
\end{figure}

\begin{figure}
\begin{center}
\epsfig{file=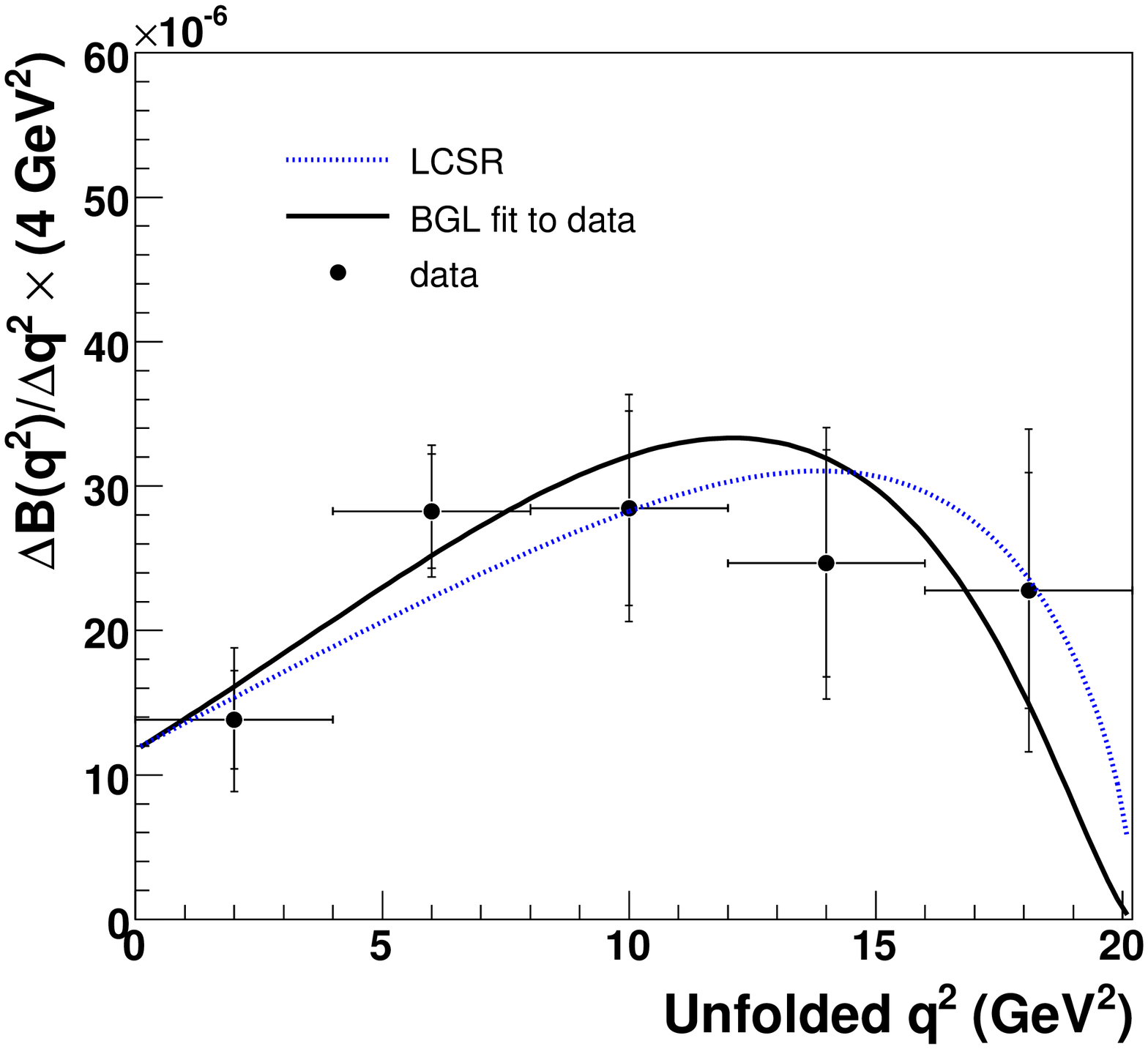,height=8.0cm}
\caption[]{\label{figFplusOmega} 
(color online) Partial \bfpilnuqq\ spectrum in 5 bins of \qq\ for \omegalnu\ 
decays. The data points are placed in the middle of each bin whose width is 
defined in Table~\ref{omegaetacerror}. The smaller error bars are statistical 
only while the larger ones also include systematic uncertainties. The data are 
also compared to a LCSR calculation~\cite{Ball05}.}  
\end{center}
\end{figure}

\begin{figure}[!htb]
\begin{center}
\epsfig{file=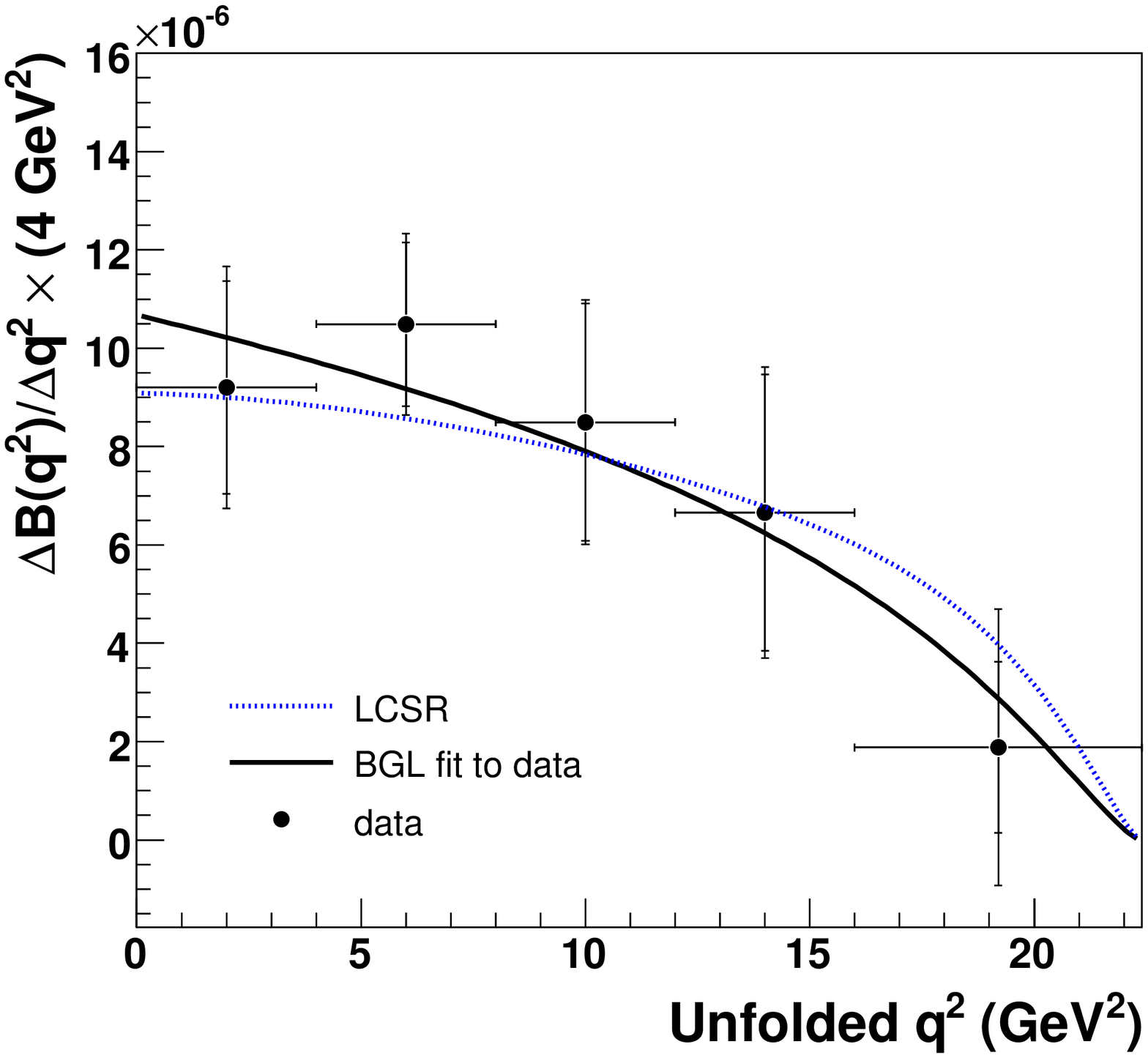,height=8.0cm}
\caption[]{\label{figFplusEta} 
(color online) Partial \bfpilnuqq\ spectrum in 5 bins of \qq\ for \etalnu\ 
decays. The data points are placed in the middle of each bin whose width is 
defined in Table~\ref{etaerror}. The smaller error bars are statistical only 
while the larger ones also include systematic uncertainties. The data are also
compared to a LCSR calculation~\cite{singlet}.}  
\end{center}
\end{figure}

 The experimental \bfpilnuqq\ distributions are displayed in 
Fig.~\ref{figFplusPi+Pi0} for the \pilnu\ decays and for the \pizlnu\ 
decays, where each point in the \pizlnu\ distribution has been normalized 
assuming isospin symmetry to hold. The two distributions are compatible. We 
show the \bfpilnuqq\ distributions in Fig.~\ref{figFplus} for the combined 
\piclnu\ decays, in Fig.~\ref{figFplusOmega} for the \omegalnu\ decays, and in 
Fig.~\ref{figFplusEta} for the \etalnu\ decays, together with theoretical 
predictions. To allow a direct comparison with the theoretical predictions, 
which do not include FSR effects, the experimental distributions in these 
figures have been obtained with the efficiency ``without FSR''. This efficiency
is given by the ratio of the total number of unfolded signal events remaining 
after all the cuts, from a simulation which includes FSR, to the total number 
of events before any cut, generated with a simulation with no FSR effects, 
{\it i.e.}, with PHOTOS switched off. 

 We obtain the \fplus\ shape from a fit to these \bfpilnuqq\ distributions. For
all decays, the $\chi^2$ function minimized in the fit to the \fplus\ shape 
uses the BGL parametrization~\cite{BGL}. Only the $\pi\ell\nu$ decays have a 
sufficient number of events to warrant the use of a two-parameter polynomial 
expansion where values of $|V_{ub}f_{+}(0)|$ can be obtained from the fit 
extrapolated to \qq\ $= 0$. For $\omega\ell\nu$ and $\eta\ell\nu$ decays we 
only use a one-parameter expansion. The resulting values of the fits are given 
in Table~\ref{BGL}. The values of $|V_{ub}f_{+}(0)|$ can be used to predict 
rates of other decays such as $B \rightarrow \pi \pi$~\cite{f0}. 

 We should note that the values of the BGL expansion parameters obtained in 
this work ($a_1/a_0=-0.92\pm0.20$, $a_2/a_0=-5.45\pm1.01$) differ somewhat from
those obtained in Ref.~\cite{Simard} ($a_1/a_0=-0.79\pm0.20$, $a_2/a_0=-4.4\pm
1.20$). Repeating the complete analysis with this new parametrization for the 
form-factor shape of the \piclnu\ decays results in only a slight change in 
\bfpiZlnu, going from $0.779\pm0.044$ to $0.773\pm0.044$, and no change in
\bfpilnu\ and \bfpiclnu. The values of $a_k/a_0$ obtained after this iteration
are given in part a) of Table~\ref{BGL}.


\begin{table*}[h]
\caption[]{\label{BGL}Fitted parameter values of the BGL parametrization for 
the exclusive semileptonic decays investigated in the present work. a) 
experimental data points only, fit parameters: $a_0$, $a_1$, $a_2$ (see Sect. 
VI); b) combined theoretical and experimental points, fit parameters: $a_0$, 
$a_1$, $a_2$, $|V_{ub}|$ (see Sect. VII).}
\begin{center}
\begin{tabular}{lccccc}
\hline\hline
 Decay mode  & $a_1/a_0$ & $a_2/a_0$ & $\chi^2$/ndf & Prob. (\%) & \VubFpz\ $\times 10^{4}$ \\ \hline
 a) \pilnu\  & -1.15 $\pm$ 0.19 & -4.52 $\pm$ 1.03 & 9.08/9 & 43.0 &\VubFpzVala\  \\
 a) \pizlnu\ & -0.63 $\pm$ 0.30 & -5.80 $\pm$ 1.24 & 3.26/8 & 91.7 &\VubFpzValza\ \\
 a) \piclnu\ & -0.93 $\pm$ 0.19 & -5.40 $\pm$ 1.00 & 4.07/9 & 90.7 &\VubFpzValca\ \\
\hline
 b) \pilnu\  & -1.25 $\pm$ 0.20 & -3.93 $\pm$ 1.19 & 9.24/12 & 68.2 &\VubFpzValb\  \\
 b) \pizlnu\ & -1.07 $\pm$ 0.28 & -3.44 $\pm$ 1.46 & 4.13/11 & 96.6 &\VubFpzValzb\ \\
 b) \piclnu\ & -1.10 $\pm$ 0.20 & -4.39 $\pm$ 1.11 & 4.58/12 & 97.1 &\VubFpzValcb\ \\
\hline
 \omegalnu\ & -5.98 $\pm$ 0.78 &         -       & 1.54/3 & 67.3 &  -       \\
 \etalnu\   & -1.71 $\pm$ 0.87 &         -       & 0.88/3 & 83.1 &  -       \\ 
\hline\hline
\end{tabular}
\end{center}
\end{table*}

\begin{table*}
\caption[]{\label{vubtable} Values of \vub\ derived from the form-factor 
calculations (first three rows) and from the value of \VubFpz\ (fourth row) for
the combined \piclnu\ decays. Value of \vub\ derived from the form-factor 
calculations (last row) for the \omegalnu\ decays. The three uncertainties on 
$|V_{ub}|$ are statistical, systematic and theoretical, respectively. 
(see Sect. VII)}
\begin{center}
\begin{tabular}{lcccccc}
\hline\hline
         & $q^2$ (\gevsq) &$\DBR$ ($10^{-4}$) &$\Delta\zeta$ (ps$^{-1}$) & $|V_{ub}|$ ($10^{-3}$) & $\chi^2$/ndf & $Prob(\chi^2)$ \\ \hline
\piclnu\                  &           &                        &                &                                          &       &      \\     
HPQCD~\cite{HPQCD06}      & $16-26.4$ & $0.37\pm 0.02\pm 0.02$ & $2.02\pm 0.55$ & $3.47\pm 0.10\pm 0.08{}^{+0.60}_{-0.39}$ & 2.7/4 & 60.1\%\\
FNAL~\cite{FNAL}          & $16-26.4$ & $0.37\pm 0.02\pm 0.02$ & $2.21{}^{+0.47}_{-0.42}$ & $3.31\pm 0.09\pm 0.07{}^{+0.37}_{-0.30}$ & 3.9/4 & 41.5\%\\
LCSR~\cite{LCSR2}         & $0-12$    & $0.83\pm 0.03\pm 0.04$ & $4.59{}^{+1.00}_{-0.85}$ & $3.46\pm 0.06\pm 0.08{}^{+0.37}_{-0.32}$ & 8.0/6 & 24.0\%\\
LCSR2~\cite{LCSR3} & $0$  &                        &                          & $3.34\pm 0.10\pm 0.05{}^{+0.29}_{-0.26}$ &       &        \\ \hline
\omegalnu\                &           &                        &                &                                          &       &        \\
LCSR3~\cite{Ball05}       & $0-20.2$  & $1.19\pm 0.16\pm 0.09$ & $14.2\pm 3.3 $ & $3.20\pm 0.21\pm 0.12{}^{+0.45}_{-0.32}$ & 2.24/5& 81.5\% \\ 
\hline\hline
\end{tabular}
\end{center}
\end{table*}

 The \qq\ distribution extracted from our data is compared in 
Fig.~\ref{figFplus} to the shape of the form factors obtained from the three
theoretical calculations listed in Table~\ref{vubtable}: the one based on Light
Cone Sum Rules~\cite{LCSR2} for $q^2 < 12$ \gevsq, and the two based on 
unquenched LQCD~\cite{HPQCD06, FNAL} for $q^2 > 16$ \gevsq. We first normalize 
the form-factor predictions to the experimental data by requiring the
integrals of both to be the same over the \qqr\ ranges of validity given in 
Table \ref{vubtable} for each theoretical prediction. Considering only 
experimental uncertainties, we then calculate the $\chi^2$ probabilities 
relative to the binned data result for various theoretical predictions in their
ranges of validity. These are given in Table~\ref{vubtable} for the combined 
\piclnu\ decays. All three calculations are compatible with the data. It should
be noted that the theoretical curves in Fig.~\ref{figFplus} have been 
extrapolated over the full \qqr\ range based on the BGL parametrization 
obtained over their \qqr\ ranges of validity. These extended ranges are only 
meant to illustrate a possible extension of the present theoretical 
calculations. As shown in Figs.~\ref{figFplusOmega} and \ref{figFplusEta}, LCSR
calculations~\cite{Ball05} and~\cite{singlet} are compatible with the data for 
the \omegalnu\ and \etalnu\ decays, respectively.

\begin{figure}[!htb]
\begin{center}
\epsfig{file=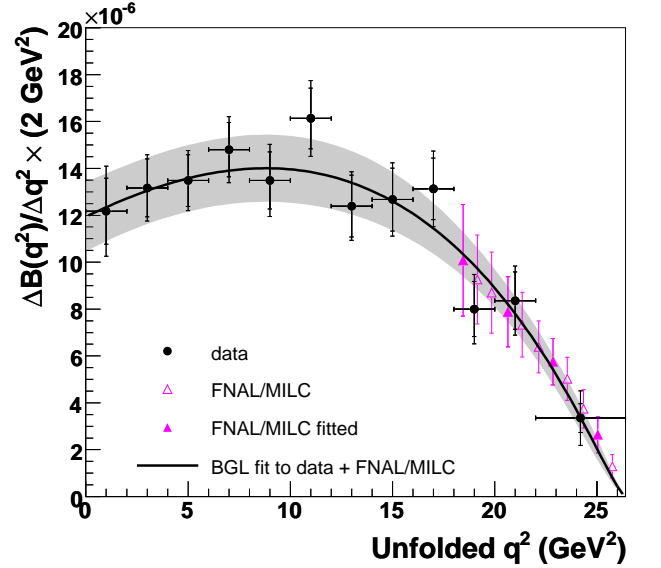,height=8.1cm}
\caption[]{\label{figLatExp} 
(color online) Simultaneous fit of the BGL parametrization~\cite{BGL} to our
experimental data (black solid points) and to four of the points of the 
FNAL/MILC predictions~\cite{FNAL} (magenta full triangles) for the \piclnu\ 
decays. The shaded band shows the uncertainty of the fitted function. The 
remaining points of the FNAL/MILC predictions (magenta empty triangles) are not
used in the fit.} 
\end{center}
\end{figure}

\section{Determination of $|V_{ub}|$}

  The magnitude of the CKM matrix element \vub\ is determined using two 
different approaches~\cite{FNAL,Jochen}.

 With the first method, we extract a value of \vub\ from the 
combined \piclnu\ \bfpilnuqq\ distributions using the relation: 
$$
|V_{ub}| = \sqrt{\Delta{\cal B}/(\tau_{B^0}\Delta \zeta)},
$$ 
where $\tau_{B^0} = 1.519 \pm 0.007$ ps~\cite{PDG10} is the $B^0$ lifetime and 
$\Delta\zeta = \Gamma/|V_{ub}|^2$ is the normalized partial decay rate 
predicted using the form-factor calculations~\cite{HPQCD06, FNAL, LCSR2}. The
quantities $\Delta{\cal B}$ and $\Delta \zeta$ are restricted to the \qqr\ 
ranges of validity given in Table \ref{vubtable}. The values of $\Delta \zeta$ 
are independent of experimental data. The values of \vub\ given in 
Table~\ref{vubtable} range from $(3.3-3.5)\times 10^{-3}$. These values are in 
good agreement with the one obtained (Table~\ref{vubtable}) from the value of 
\VubFpz\ $=(8.7\pm 0.3)\times 10^{-4}$ measured in this work, using the value 
of $f_+(0) = 0.26{}^{+0.020}_{-0.023}$ determined in a recent LCSR 
calculation~\cite{LCSR3}. They are also compatible with the value of \vub\ 
determined from the \omegalnu\ data, as shown in Table \ref{vubtable}. A value
of \vub\ is not extracted from the \etalnu\ decays because the theoretical 
partial decay rate is not sufficiently precise for these decays. 

  With the second method, we perform a simultaneous fit to the most recent
lattice results~\cite{FNAL} and our present experimental data to take advantage
of all the available information on the form factor from the data (shape) and 
theory (shape and normalization). 

  The $\chi^2$ function for the simultaneous fit is written as:
\begin{eqnarray*}
\chi^2 &=& \chi^2(data) + \chi^2(lattice) \nonumber \\
       &=& \sum^{n_{bins}}_{i,j=1}\Delta^{data}_i
(V^{data}_{ij})^{-1}\Delta^{data}_{j} + 
\sum^{n_{points}}_{\ell,m=1}\Delta^{lat}_{\ell}
(V^{lat}_{\ell m})^{-1}\Delta^{lat}_{m} \nonumber
\end{eqnarray*}
\noindent 
where:
\begin{eqnarray*}
\Delta^{data}_{i}&=& \left (\frac{\Delta{\cal B}}{\Delta q^2}\right)^{data}_{i}
\nonumber \\
                  & & - \frac{|V_{ub}|^2}{\Delta q^2_i}\int_{\Delta q^2_i}
\frac{\tau_{B^0}G^2_F}{24\pi^3}p^3_{\pi}(q^2)|f_+(q^2;\alpha)|^2dq^2 \nonumber 
\\
\Delta^{lat}_{\ell} &=& \frac{G^2_F}{24\pi^3}p^3_{\pi}(q^2_{\ell})
\{|f^{lat}_+(q^2_{\ell})|^2 - |f_+(q^2_{\ell};\alpha)|^2\} \nonumber
\end{eqnarray*}
where $G_F$ is the Fermi constant, $\alpha$ denotes the set of parameters for a
chosen parametrization of $f_+(q^2)$, 
$\left (\Delta{\cal B}/\Delta q^2\right)^{data}_{i}$ is the 
measured partial BF \qq\ spectrum, $|f^{lat}_+(q^2_{\ell})|$ are the LQCD 
form-factor predictions, $q^2_{\ell}$ is the value of $q^2$ for which we have a
theoretical point, and $(V^{data}_{ij})^{-1}$ and $(V^{lat}_{\ell m})^{-1}$ are
the inverse covariance matrices for data and theory, respectively. In our work,
the function $|f_+(q^2_{\ell};\alpha)|$ contains the coefficients $a_k$ of the 
BGL parametrization. The result of the simultaneous fit for \piclnu\ decays is 
shown in Fig.~\ref{figLatExp}, where with four theoretical points, we obtain 
the values of the BGL parametrization given in Table~\ref{BGL} and 
$a_0 = (2.26\pm 0.20)\times 10^{-2}$. The two values of $a_k/a_0$ are very 
similar to those obtained from a fit to the experimental data alone using the 
BGL parametrization. This is not surprising since the data dominate the fit. We
have only used the subset with four of the 12 theoretical points in our 
simultaneous fit since adjacent points are very strongly 
correlated~\cite{FNAL}. Alternative choices of subset give compatible results.
The results shown for the \pilnu\ and \pizlnu\ decays in Table~\ref{BGL} are 
consistent with those obtained for the combined \piclnu\ decays. 

 The fit also yields: $|V_{ub}|= (3.25\pm 0.31)\times 10^{-3}$. The previous 
\babar\ result~\cite{Jochen} of $|V_{ub}|=(2.95\pm 0.31) \times 10^{-3}$ is 
about 1 standard deviation smaller. This fairly large difference can be 
understood from the fact the determination of \vub\ from the combined data-LQCD
fit is most sensitive to the points at high $q^2$, where the changes due to the
improved hybrid treatment leads to differences larger than those expected on 
the basis of the variation in the total BF value. The present value of \vub\ 
supersedes the one from Ref.~\cite{Jochen}.    

 Since the total uncertainty of 9.5\% on the value of \vub\ results from the 
simultaneous fit to data and LQCD predictions, it is not so easy to identify 
the contributions from experiment and theory to this uncertainty. We estimate 
that the total uncertainty of 4.9\% in the BF measurement is equivalent to an
experimental uncertainty of 2.4\% in the value of \vub. The contribution to the
uncertainty from the shape of the \qqr\ spectrum is determined by varying the 
fit  parameters $a_1/a_0$ and $a_2/a_0$ within their uncertainties, and taking
into account their correlation. This yields a contribution of 3.1\% to the 
uncertainty in the value of \vub. The remaining uncertainty of 8.7\% arises 
from the form-factor normalization provided by theory.

\section{Summary}

 In summary, we have measured the partial BFs of \pilnu\ and combined \piclnu\ 
decays in 12 bins of \qq, of \pizlnu\ decays in 11 bins of \qq, and of 
\omegalnu\ and \etalnu\ decays in five bins of \qq. From the \piclnu\ 
distributions, we extract the \fplus\ shapes that are found to be compatible,
in the appropriate \qq\ range, with all three theoretical predictions 
considered for these decays. LCSR calculations are also found to be consistent 
with our measured \omegalnu~\cite{Ball05} and \etalnu~\cite{singlet} 
\bfpilnuqq\ distributions. The BGL parametrization fits our \pilnu, \pizlnu\ 
and \piclnu\ data well and allows us to obtain the value of $|V_{ub}f_+(0)|$. 
Our measured branching fractions of the five decays reported in this work lead 
to some improvement in our knowledge of the composition of the inclusive 
charmless semileptonic decay rate. In particular, the form-factor shapes are 
now better defined, especially for the $\pi\ell\nu$ decays. Our values of the 
total BF for \etaetaplnu\ decays are in good agreement with our earlier 
results~\cite{Simard} and supersede them. The value of the ratio 
\bfetaplnu/\bfetalnu\ = $0.63 \pm0.24_{stat} \pm0.11_{syst}$ allows a 
significant gluonic singlet contribution to the $\eta^{\prime}$ form 
factor~\cite{singlet, singlet2}. In spite of large differences in the analysis 
methods for the \omegalnu\ decays, our total BF is in good agreement with our 
previous result~\cite{Wulsin}. The present precise value of the total BF for 
\piclnu\ decays is slightly larger than the most recent \babar\ 
results~\cite{Jochen,Simard} for the reasons expounded in Sect. VI. It 
supersedes both results. It is in good agreement with the recent Belle result.
Our value has comparable precision to the present world average~\cite{PDG10}. 
For \piclnu\ decays, we obtain values of \vub\ for three different QCD 
calculations. The results are in good agreement with those of 
Refs.~\cite{Jochen,Simard}. The three values are compatible with the value of 
\vub\ obtained from our measured value of \VubFpz, with our value of \vub\ 
extracted from the \omegalnu\ data, and with the value of $|V_{ub}|
=(3.25\pm 0.31)\times 10^{-3}$ determined from the simultaneous fit to our 
experimental data and the LQCD theoretical predictions. It is compatible with 
the Belle result~\cite{Belle} of $|V_{ub}|=(3.43\pm 0.33)\times 10^{-3}$. The 
tension between our values of $|V_{ub}|$ and the value of 
$|V_{ub}|=(4.27\pm 0.38)\times 10^{-3}$ \cite{PDG10} measured in inclusive 
semileptonic $B$ decays remains significant.

\section{Acknowledgments}
 We are grateful for the extraordinary contributions of our 
\pep2\ colleagues in achieving the excellent luminosity and machine conditions 
that have made this work possible. The success of this project also relies 
critically on the 
expertise and dedication of the computing organizations that support \babar.
The collaborating institutions wish to thank SLAC for its support and the kind 
hospitality extended to them. This work is supported by the US Department of 
Energy and National Science Foundation, the Natural Sciences and Engineering 
Research Council (Canada), the Commissariat \`a l'Energie Atomique and Institut
National de Physique Nucl\'eaire et de Physique des Particules (France), the
Bundesministerium f\"ur Bildung und Forschung and Deutsche 
Forschungsgemeinschaft (Germany), the Istituto Nazionale di Fisica Nucleare 
(Italy), the Foundation for Fundamental Research on Matter (The Netherlands),
the Research Council of Norway, the Ministry of Education and Science of the 
Russian Federation, Ministerio de Ciencia e Innovaci\'on (Spain), and the
Science and Technology Facilities Council (United Kingdom). Individuals have 
received support from the Marie-Curie IEF program (European Union), the A. P. 
Sloan Foundation (USA) and the Binational Science Foundation (USA-Israel).

\vspace{1cm}
\section{Appendix}

  In Tables~\ref{cutSummaryPic}-\ref{cutSummary}, we give the functions 
describing the \qq\ dependence of the selections used to reduce the backgrounds
in the five decays under study. In Table~\ref{ScalingF} we give the 
values of the scaling factors obtained in our fit to the data for each decay 
channel. In Tables~\ref{table1}-\ref{table8}, we present the full correlation 
matrices (elements in \%) of the fitting scaling factors for all the decay 
channels under investigation.

 The list of all the systematic uncertainties, as well as their values for the 
partial and total BFs, are given in Tables~\ref{picerror},~\ref{pierror},
~\ref{pizerror},~\ref{omegaetacerror} and~\ref{etaerror} for the five decays. 
In Table~\ref{picerror}, we have one column for each bin of \qqr\ and three 
columns for various ranges of \qqr\ as well as the last column for the global 
result. In row 1, ``Fitted yield'', we give the raw fitted yield as the number 
of events. In row 2, ``Yield statistical error'', we give the statistical 
uncertainty in \% for each fitted yield. In row 3, ``Unfolded yield'', we give 
the yields from row 1 unfolded to give the true values of the yields in each 
bin, expressed as the number of events. In rows 4 and 6, ``Efficiency'', we 
give the efficiency in \% attached to each yield. In rows 5 and 7, ``Eff. 
(without FSR)'', we give the efficiency in \%, modified to remove the FSR 
effect. In row 8, ``\DBR", we give the values of the partial BFs computed as 
usual using the true (unfolded) yields and the efficiencies with FSR. In row 9,
``\DBR\ (without FSR)", we give the values of the partial BFs computed as usual
using the true (unfolded) yields and the efficiencies modified to remove the 
FSR effect. In rows 10 - 42, we give the contributions in \% to the relative 
systematic uncertainties for each value of \DBR\ as a function of \qqr. In row 
43, ``Signal MC statistical error'', we give the statistical uncertainty due to
the number of MC signal events. In row 44, ``Total systematic error'', we give 
the total systematic uncertainty in \% for each value of \DBR, obtained as the 
sum in quadrature of all the systematic uncertainties in each column. In row 
45, ``Fit error'' (also denoted total statistical error), we give the statistical uncertainty in \% for 
each value of \DBR\ obtained from propagating the statistical uncertainties on 
the raw fitted yields, following the unfolding process and taking into account 
the efficiencies. In row 46, ``Total error'', we first give the total 
uncertainty in \% for each value of \DBR, obtained as the sum in quadrature of 
the total systematic error and the fit error. We then give, in 
the last four columns, the total uncertainties in \% for each range of \qqr, 
obtained as the sum in quadrature of the total errors for the appropriate 
number of \qqr\ bins. A similar description applies to the other tables. 

\begin{figure*}
\begin{center}
\epsfig{file=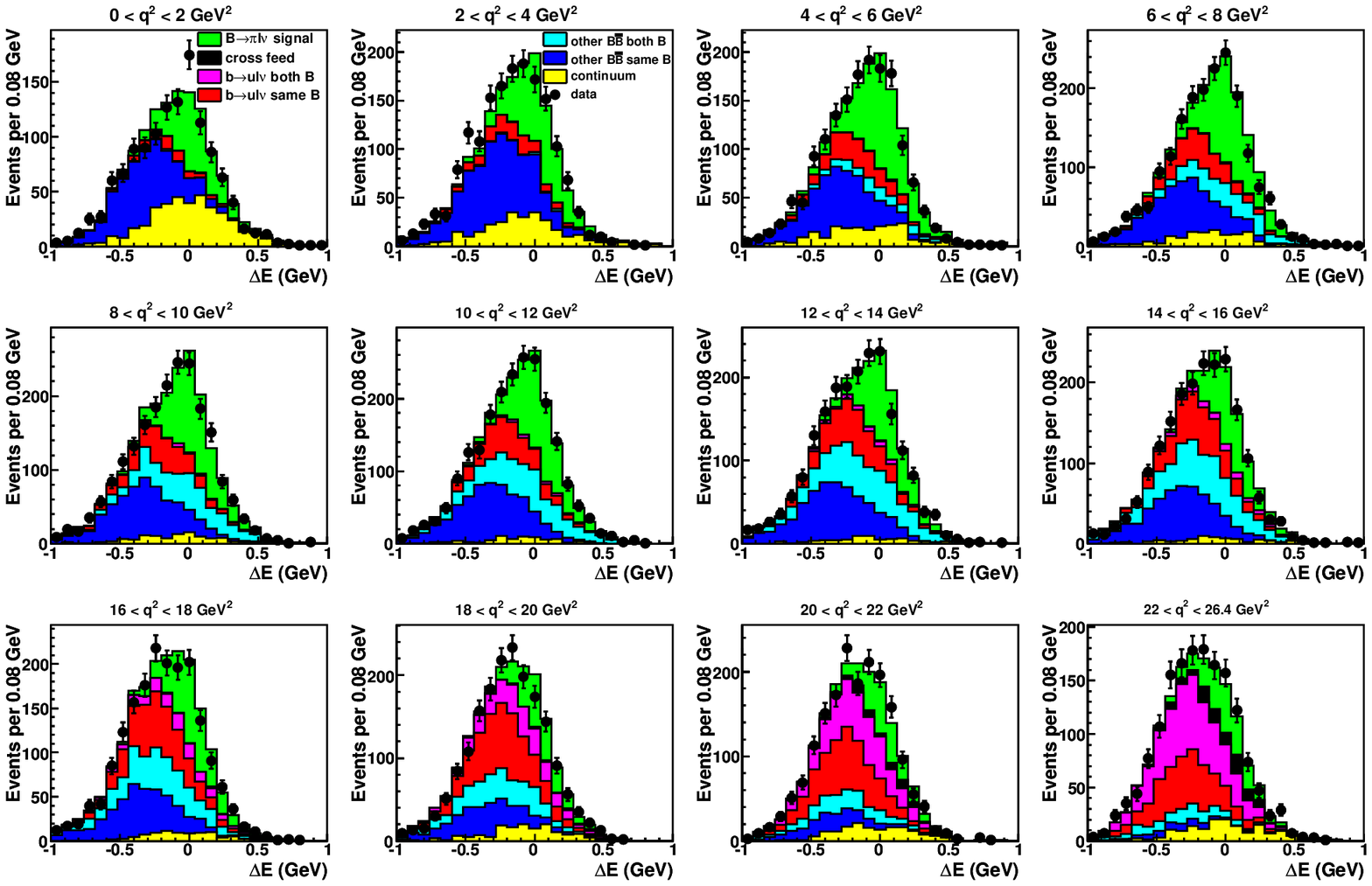,height=10cm}
\caption[]
{\label{dEFitProjDataPic} 
(color online) \DeltaE\ yield fit projections in the signal-enhanced region, 
with \mes\ $>$ 5.2675 \gev, obtained in 12~\qq~bins from the fit to the 
experimental data for combined \pilnu\ and \pizlnu\ decays. The fit was done 
using the full \DeltaE-\mes fit region.}
\end{center}
\end{figure*}

\begin{figure*}
\begin{center}
\epsfig{file=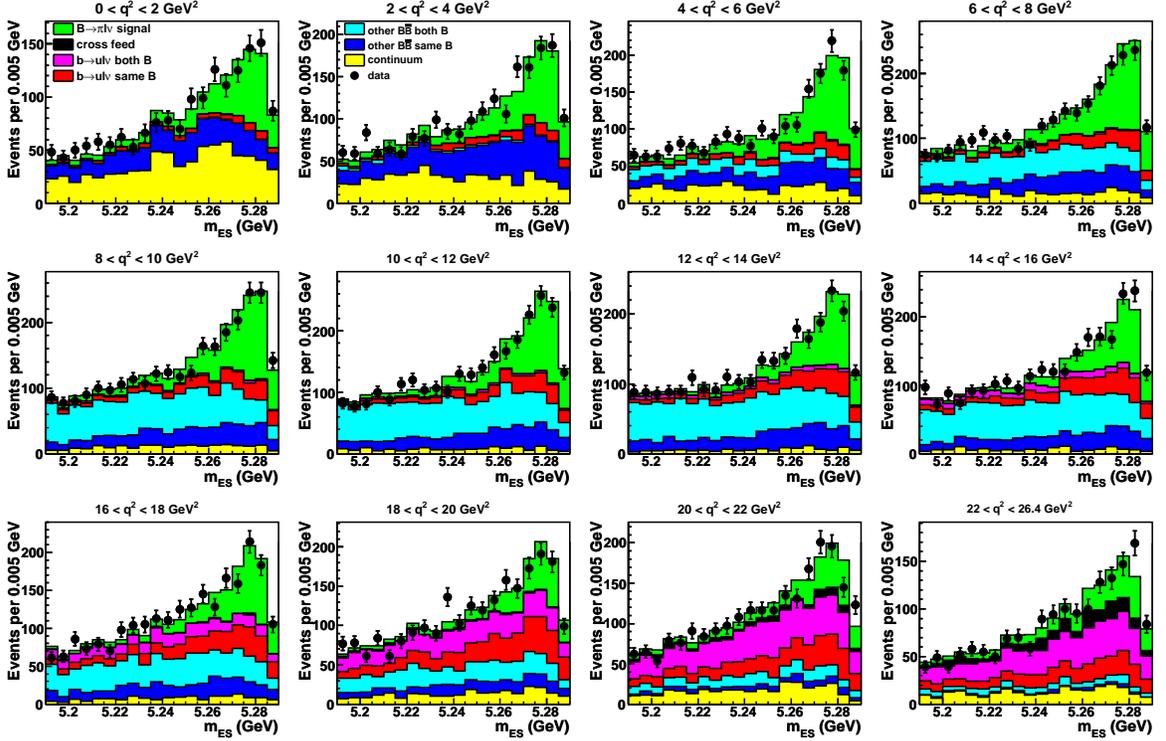,height=10cm}
\caption[]
{\label{mESFitProjDataPic} 
(color online) \mes\ yield fit projections in the signal-enhanced region, with 
$-0.16<\Delta E<0.20~\gev$, obtained in 12~\qq~bins from the fit to the 
experimental data for combined \pilnu\ and \pizlnu\ decays. The fit was done 
using the full \DeltaE-\mes\ fit region.}
\end{center}
\end{figure*}

 In our analysis, we compute the covariance matrix for each source of 
uncertainty, and use these matrices to calculate the uncertainties on the total
BFs. The correlation matrices for the total statistical and systematic 
uncertainties are given in Tables~\ref{StatCovPic} and~\ref{SystCovPic} for the
combined \piclnu\ yields, in Tables~\ref{StatCovPi} and~\ref{SystCovPi} for 
the \pizlnu\ yields, in Table~\ref{CovOmega} for the \omegalnu\ yields and in 
Table~\ref{CovEta} for the \etalnu\ yields. Finally, detailed \DeltaE\ and 
\mes\ fit projections in each \qqr\ bin are also shown in 
Figs.~\ref{dEFitProjDataPic} and~\ref{mESFitProjDataPic}, respectively, for the
combined \piclnu\ decays.

\begin{table*}
\begin{center}
\caption[]{
\label{cutSummaryPic} \qq-dependent selections used in \pilnu\ decays.}
\begin{tabular}{l}
\hline\hline
$\cos\theta_{\ell} < 0.85$ for all values of \qqr\ \\
$\cos\theta_{\ell} > 0.00000352q^{10}-0.000235q^8+0.00513q^6-0.0383q^4
+0.0299q^2-0.315$ \\
$m^2_{miss}/2E_{miss} > -0.5$~\gev\ for all values of \qqr\ \\
$m^2_{miss}/2E_{miss} < -0.0000499q^8+0.00238q^6-0.0342q^4+0.129q^2+
0.895$~\gev\ \\
$\cos\theta_{thrust} < -0.00000578q^8+0.000319q^6-0.00737q^4+0.0807q^2
+0.551$ \\
$\theta_{miss} > 0.000829q^4-0.0125q^2+0.34$ rad \\
$L2 < -0.000147q^6+0.00141q^4+0.0579q^2+1.54$ \\
(\qqr\ is given in units of \gevsq) \\
\hline\hline
\end{tabular}
\end{center}
\end{table*}

\begin{table*}
\begin{center}
\caption[]{
\label{cutSummaryPiz} \qq-dependent selections used in \pizlnu\ decays.}
\begin{tabular}{l}
\hline\hline
$\cos\theta_{\ell} < 0.9$ for all values of \qqr\ \\
$\cos\theta_{\ell} > -0.75$, \qqr\ $<=$ 1.5 \gevsq\ \\
$\cos\theta_{\ell} > 0.00000891q^{10}-0.00057q^8+0.0128q^6-0.12q^4
+0.456q^2-1.18$, \qqr\ $>$ 1.5 \gevsq\ \\
$m^2_{miss}/2E_{miss} > -0.4$~\gev\ for all values of \qqr\ \\
$m^2_{miss}/2E_{miss} < 0.000000167q^{10}-0.0000448q^8+0.00201q^6
-0.0315q^4+0.152q^2+0.744$~\gev\ \\
$\cos\theta_{thrust} < -0.00000146q^{10}+0.0000679q^8-0.000816q^6
-0.00298q^4+0.0991q^2+0.431$ \\
$\theta_{miss} <2.9$ rad for all values of \qqr\ \\
$\theta_{miss} > 0.00000464q^{10}-0.000252q^8+0.00474q^6-0.0357q^4
+0.0996q^2+0.306$ rad \\
$L2 < -0.00000399q^{10}+0.000199q^8-0.00315q^6+0.0127q^4+0.0883q^2+1.3$ \\
$p^{_*}_{lep} > 0.00000398q^{10}-0.000251q^8+0.00538q^6-0.0459q^4
+0.233q^2+0.29$ \gev\ \\
(\qqr\ is given in units of \gevsq) \\
\hline\hline
\end{tabular}
\end{center}
\end{table*}

\begin{table*}
\begin{center}
\caption[]{
\label{cutSummaryOmega} \qq-dependent selections used in \omegalnu\ \\
decays.}
\begin{tabular}{l}
\hline\hline
$m^2_{miss}/2E_{miss} > -0.4$~\gev\ for all values of \qqr\ \\
$m^2_{miss}/2E_{miss} < -0.00000393q^{10}+0.0000411q^8+0.00305q^6
-0.0623q^4+0.326q^2+0.49$~\gev\ \\
$\theta_{miss} <2.65$ rad for all values of \qqr\ \\
$\theta_{miss} > -0.00000244q^{10}+0.000167q^8-0.00411q^6+0.0434q^4
-0.173q^2+0.483$ rad \\
$L2 < -0.00000659q^{10}+0.000316q^8-0.00548q^6+0.0381q^4-0.0458q^2+1.58$\\
$p^{_*}_{lep} > -0.0000139q^{10}+0.000929q^8-0.0228q^6+0.239q^4-0.838q^2
+1.43$~\gev\ \\
$p^{_*}_{lep}+p^{_*}_{\omega} > -0.00000581q^{10}+0.000449q^8-0.0129q^6+0.161q^4
-0.801q^2+4.28$~\gev\ \\
$p^{_*}_{\pi^0} > -0.00000479q^{10}+0.000292q^8-0.00651q^6+0.0641q^4-0.278q^2
+0.787$~\gev\ \\
(\qqr\ is given in units of \gevsq) \\
\hline\hline
\end{tabular}
\end{center}
\end{table*}

\begin{table*}
\begin{center}
\caption[]{
\label{cutSummaryEta} \qq-dependent selections used in \etalnu\ ($\eta 
\rightarrow \gamma\gamma$) decays.}
\begin{tabular}{l}
\hline\hline
$\cos\theta_{\ell} < 0.9$ for all values of \qqr\ \\
$\cos\theta_{\ell} > -0.0000582q^8+0.00242q^6-0.0302q^4+0.115q^2-0.793$ \\
$m^2_{miss}/2E_{miss} > -0.4$~\gev\ for all values of \qqr\ \\
$m^2_{miss}/2E_{miss} < 0.00000775q^{10}-0.000579q^8+0.0152q^6-0.171q^4
+0.72q^2+0.168$~\gev\ \\
$\cos\theta_{thrust} < -0.0000332q^8+0.00153q^6-0.0249q^4+0.168q^2+0.42$ \\
$\theta_{miss} < 2.8$ rad for all values of \qqr\ \\
$\theta_{miss} > 0.0000782q^8-0.00363q^6+0.0567q^4-0.32q^2+0.863$ rad \\
$p^{_*}_{miss} < 3.2625$~\gev\ \\
$p^{_*}_{miss} > 0.000101q^8-0.00456q^6+0.0661q^4-0.239q^2+0.819$~\gev\ \\
(\qqr\ is given in units of \gevsq) \\
\hline\hline
\end{tabular}
\end{center}
\end{table*}

\begin{table*}
\begin{center}
\caption[]{
\label{cutSummaryEta2} \qq-dependent selections used in \etalnu\ ($\eta 
\rightarrow \pi^+\pi^-\pi^0$) decays.}
\begin{tabular}{l}
\hline\hline
$m^2_{miss}/2E_{miss} > -0.4$~\gev\ for all values of \qqr\ \\
$m^2_{miss}/2E_{miss} < 0.0000082q^{10}-0.000551q^8+0.0136q^6-0.15q^4
+0.655q^2+0.0359$~\gev\ \\
$\cos\theta_{thrust} <-0.0000235q^8+0.00087q^6-0.0126q^4+0.0831q^2+0.629$\\
$\theta_{miss} < 2.85$ rad for all values of \qqr\ \\
$\theta_{miss} > 0.0000682q^8-0.00298q^6+0.0425q^4-0.206q^2+0.595$ rad \\
(\qqr\ is given in units of \gevsq) \\
\hline\hline
\end{tabular}
\end{center}
\end{table*}

\begin{table*}
\begin{center}
\caption[]{
\label{cutSummary} \qq-dependent selections used in \etaplnu\ decays.}
\begin{tabular}{l}
\hline\hline
$m^2_{miss}/2E_{miss} > -0.3$~\gev\ for all values of \qqr\ \\
$m^2_{miss}/2E_{miss} < 0.35q^2+0.325$~\gev, \qqr\ $<$ 2.5 \gevsq\ \\
$m^2_{miss}/2E_{miss} < 1.2$~\gev, 2.5 $<$ \qqr\ $<$ 4.5 \gevsq \\
$m^2_{miss}/2E_{miss} < -0.1q^2+1.65$~\gev, \qqr\ $>$ 4.5 \gevsq\ \\
$\cos\theta_{thrust} < 0.05q^2+0.575$, \qqr\ $<$ 6.5 \gevsq\ \\
$\cos\theta_{thrust} < 0.9$, 6.5 $<$ \qqr\ $<$ 12.5 \gevsq\ \\
$\cos\theta_{thrust} < -0.05q^2+1.525$, \qqr\ $>$ 12.5 \gevsq\ \\
$\theta_{miss} > -0.1q^2+0.45$ rad, \qqr\ $<$ 2.5 \gevsq\ \\
$\theta_{miss} > 0.2$ rad, 2.5 $<$ \qqr\ $<$ 5.5 \gevsq\ \\
$\theta_{miss} > 0.05q^2-0.075$ rad, \qqr\ $>$ 5.5 \gevsq\ \\
(\qqr\ is given in units of \gevsq) \\
\hline\hline
\end{tabular}
\end{center}
\end{table*}

\begin{table*}
\caption[]{\label{ScalingF} Values of the scaling factors given by the fit 
results for each decay channel. The superscripts ($u\ell\nu,1$) and 
($u\ell\nu,2$) represent the \ulnu\ same-$B$ and both-$B$ backgrounds, 
respectively, and likewise for the \obb\ background.}
\begin{center}
\begin{tabular}{lcccccccc}
\hline\hline
&$\pi\ell^+\nu$&$\pi^-\ell^+\nu$&$\pi^0\ell^+\nu$&$\omega\ell^+\nu$&$\eta\ell^+\nu$
&$\eta\ell^+\nu~(\gamma\gamma)$&$\eta\ell^+\nu~(3\pi)$&$\eta^{\prime}\ell^+\nu$\\ \hline
$p^{signal}_{1}$ & $0.88\pm0.09$&$0.89\pm0.11$&$0.93\pm0.14$&$0.98\pm0.20$&$1.08\pm0.23$&$1.20\pm0.27$&$1.29\pm0.29$&$1.01\pm0.35$\\
$p^{signal}_{2}$ & $0.92\pm0.06$&$0.89\pm0.07$&$1.05\pm0.10$&$1.26\pm0.15$&$1.23\pm0.17$&$0.90\pm0.19$&-&-\\
$p^{signal}_{3}$ & $0.94\pm0.05$&$0.90\pm0.06$&$1.04\pm0.09$&$1.06\pm0.20$&$1.11\pm0.26$&$1.12\pm0.31$&-&-\\
$p^{signal}_{4}$ & $1.00\pm0.05$&$1.01\pm0.06$&$0.99\pm0.09$&$0.90\pm0.21$&$1.01\pm0.37$&$0.78\pm0.39$&-&-\\
$p^{signal}_{5}$ & $0.99\pm0.06$&$1.02\pm0.07$&$0.91\pm0.11$&$1.02\pm0.28$&$0.70\pm0.48$&$0.87\pm0.62$&-&-\\
$p^{signal}_{6}$ & $1.12\pm0.07$&$1.13\pm0.08$&$1.08\pm0.12$&-&-&-&-&-\\
$p^{signal}_{7}$ & $1.00\pm0.08$&$0.98\pm0.09$&$1.09\pm0.14$&-&-&-&-&-\\
\hline\hline
\end{tabular}
\end{center}
\end{table*}

\begin{table*}
\caption[]{\label{table1} Correlation matrix (elements in \%) of the fitted 
scaling factors for the \piclnu\ decay channel. The superscripts ($u\ell\nu,1$)
and ($u\ell\nu,2$) represent the \ulnu\ same-$B$ and both-$B$ backgrounds,
respectively, and likewise for the \obb\ background.}
\begin{center}
\begin{tabular}{lcccccccccccccccccccccc}
\hline\hline
\\[-11pt]
& $p^{\pi}_{1}$ &$p^{\pi}_{2}$ &$p^{\pi}_{3}$ &$p^{\pi}_{4}$ &$p^{\pi}_{5}$ &$p^{\pi}_{6}$ 
&$p^{\pi}_{7}$ &$p^{\pi}_{8}$ &$p^{\pi}_{9}$ &$p^{\pi}_{10}$ &$p^{\pi}_{11}$ &$p^{\pi}_{12}$ &$p^{u\ell\nu,1}_{1}$
&$p^{u\ell\nu,1}_{2}$&$p^{u\ell\nu,2}_{1}$&$p^{u\ell\nu,2}_{2}$&$p^{B\bar{B},1}_{1}$
&$p^{B\bar{B},1}_{2}$&$p^{B\bar{B},2}_{1}$&$p^{B\bar{B},2}_{2}$
&$p^{cont}_{1}$&$p^{cont}_{2}$ \\[+1pt]\hline
$p^{\pi}_{1}$ & 100&27&19&8&0&3&2&5&7&11&10&1&-6&-4&13&1&24&1&18&10&-60&0\\
$p^{\pi}_{2}$ & 27&100&16&9&4&2&2&4&7&7&6&1&-9&-3&7&1&4&2&15&8&-36&0\\
$p^{\pi}_{3}$ & 19&16&100&15&12&9&10&10&12&5&5&1&-26&-9&8&3&14&18&1&5&-25&1\\
$p^{\pi}_{4}$ & 8&9&15&100&19&13&14&10&12&3&2&2&-29&-10&6&4&11&28&-17&-1&-7&2\\
$p^{\pi}_{5}$ & 0&4&12&19&100&14&16&10&11&1&0&2&-30&-11&4&5&9&32&-27&-5&2&2\\
$p^{\pi}_{6}$ & 3&2&9&13&14&100&17&9&12&2&3&1&-25&-5&0&1&19&10&-15&9&-4&1\\
\hline\hline
\end{tabular}
\end{center}
\end{table*}

\begin{table*}
\caption[]{\label{table2} Correlation matrix (elements in \%) of the fitted 
scaling factors for the \pilnu\ decay channel. The superscripts ($u\ell\nu,1$) 
and ($u\ell\nu,2$) represent the \ulnu\ same-$B$ and both-$B$ backgrounds,
respectively, and likewise for the \obb\ background.}
\begin{center}
\begin{tabular}{lcccccccccccccccccccccc}
\hline\hline
\\[-11pt]
& $p^{\pi^-}_{1}$ &$p^{\pi^-}_{2}$ &$p^{\pi^-}_{3}$ &$p^{\pi^-}_{4}$ &$p^{\pi^-}_{5}$ &$p^{\pi^-}_{6}$ 
&$p^{\pi^-}_{7}$ &$p^{\pi^-}_{8}$ &$p^{\pi^-}_{9}$ &$p^{\pi^-}_{10}$ &$p^{\pi^-}_{11}$ &$p^{\pi^-}_{12}$ &$p^{u\ell\nu,1}_{1}$
&$p^{u\ell\nu,1}_{2}$&$p^{u\ell\nu,2}_{1}$&$p^{u\ell\nu,2}_{2}$&$p^{B\bar{B},1}_{1}$
&$p^{B\bar{B},1}_{2}$&$p^{B\bar{B},2}_{1}$&$p^{B\bar{B},2}_{2}$
&$p^{cont}_{1}$&$p^{cont}_{2}$ \\[+1pt] \hline
$p^{\pi^-}_{1}$ & 100&29&20&7&0&3&2&5&7&9&7&1&-6&-3&14&1&27&1&19&9&-64&1\\
$p^{\pi^-}_{2}$ & 29&100&16&7&3&2&2&4&6&5&5&0&-7&-2&8&1&4&1&15&8&-37&0\\
$p^{\pi^-}_{3}$ & 20&16&100&13&11&9&10&9&11&4&4&1&-24&-8&8&3&14&17&0&4&-26&2\\
$p^{\pi^-}_{4}$ & 7&7&13&100&18&12&14&10&11&2&3&1&-27&-10&5&4&9&27&-19&-3&-6&2\\
$p^{\pi^-}_{5}$ & 0&3&11&18&100&13&15&9&10&0&1&1&-27&-10&4&4&7&31&-28&-7&3&2\\
$p^{\pi^-}_{6}$ & 3&2&9&12&13&100&17&8&11&2&3&1&-24&-4&0&1&17&8&-15&9&-4&1\\
\hline\hline
\end{tabular}
\end{center}
\end{table*}

\begin{table*}
\caption[]{\label{table3} Correlation matrix (elements in \%) of the fitted 
scaling factors for the \pizlnu\ decay channel. The superscripts ($u\ell\nu,1$)
and ($u\ell\nu,2$) represent the \ulnu\ same-$B$ and both-$B$ backgrounds,
respectively, and likewise for the \obb\ background.}
\begin{center}
\begin{tabular}{lcccccccccccccccc}
\hline\hline
\\[-11pt]
& $p^{\pi^0}_{1}$ &$p^{\pi^0}_{2}$ &$p^{\pi^0}_{3}$ &$p^{\pi^0}_{4}$ &$p^{\pi^0}_{5}$ &$p^{\pi^0}_{6}$ 
&$p^{\pi^0}_{7}$ &$p^{\pi^0}_{8}$ &$p^{\pi^0}_{9}$ &$p^{\pi^0}_{10}$ &$p^{\pi^0}_{11}$ &$p^{u\ell\nu,1}_{1}$
&$p^{u\ell\nu,2}_{1}$&$p^{B\bar{B},1}_{1}$&$p^{B\bar{B},2}_{1}$&$p^{cont}_{1}$ \\[+1pt] \hline
$p^{\pi^0}_{1}$ & 100&17&13&9&5&2&1&2&5&7&4&-7&13&10&10&-43\\
$p^{\pi^0}_{2}$ & 17&100&13&9&6&3&3&4&6&7&3&-12&12&5&13&-32\\
$p^{\pi^0}_{3}$ & 13&13&100&10&9&7&6&7&8&8&1&-20&17&8&8&-23\\
$p^{\pi^0}_{4}$ & 9&9&10&100&11&9&8&8&8&7&-1&-21&19&11&-1&-14\\
$p^{\pi^0}_{5}$ & 5&6&9&11&100&12&11&10&8&7&-4&-26&23&17&-11&-6\\
$p^{\pi^0}_{6}$ & 2&3&7&9&12&100&10&9&7&6&-4&-22&20&16&-14&-2\\
$p^{\pi^0}_{7}$ & 1&3&6&8&11&10&100&9&8&7&-1&-25&16&14&-7&0\\
$p^{\pi^0}_{8}$ & 2&4&7&8&10&9&9&100&9&8&1&-26&13&14&0&-1\\
$p^{\pi^0}_{9}$ & 5&6&8&8&8&7&8&9&100&12&9&-27&6&15&7&-7\\
$p^{\pi^0}_{10}$ &7&7&8&7&7&6&7&8&12&100&20&-23&-5&16&9&-11\\
$p^{\pi^0}_{11}$ &4&3&1&-1&-4&-4&-1&1&9&20&100&6&-57&-4&21&-8\\
$p^{u\ell\nu,1}_{1}$&-7&-12&-20&-21&-26&-22&-25&-26&-27&-23&6&100&-55&-63&-12&10\\
$p^{u\ell\nu,2}_{1}$&13&12&17&19&23&20&16&13&6&-5&-57&-55&100&50&-25&-25\\
$p^{B\bar{B},1}_{1}$&10&5&8&11&17&16&14&14&15&16&-4&-63&50&100&-33&-35\\
$p^{B\bar{B},2}_{1}$&10&13&8&-1&-11&-14&-7&0&7&9&21&-12&-25&-33&100&-7\\
$p^{cont}_{1}$&-43&-32&-23&-14&-6&-2&0&-1&-7&-11&-8&10&-25&-35&-7&100\\
\hline\hline
\end{tabular}
\end{center}
\end{table*}

\begin{table*}
\caption[]{\label{table4} Correlation matrix (elements in \%) of the fitted 
scaling factors for the \omegalnu\ decay channel.}
\begin{center}
\begin{tabular}{lcccccccc}
\hline\hline
\\[-11pt]
& $p^{\omega}_{1}$ &$p^{\omega}_{2}$ &$p^{\omega}_{3}$ &$p^{\omega}_{4}$ &$p^{\omega}_{5}$
&$p^{u\ell\nu}_{1}$&$p^{B\bar{B}}_{1}$&$p^{cont}_{1}$ \\[+1pt] \hline
$p^{\omega}_{1}$ & 100&12&7&10&2&12&7&-46\\
$p^{\omega}_{2}$ & 12&100&11&7&11&-5&-36&7\\
$p^{\omega}_{3}$ & 7&11&100&19&26&-24&-3&0\\
$p^{\omega}_{4}$ & 10&7&19&100&58&-59&26&-15\\
$p^{\omega}_{5}$ & 2&11&26&58&100&-85&19&3\\
$p^{u\ell\nu}_{1}$&12&-5&-24&-59&-85&100&-19&-22\\
$p^{B\bar{B}}_{1}$&7&-36&-3&26&19&-19&100&-61\\
$p^{cont}_{1}$&-46&7&0&-15&3&-22&-61&100\\
\hline\hline
\end{tabular}
\end{center}
\end{table*}

\begin{table*}
\caption[]{\label{table5} Correlation matrix (elements in \%) of the fitted 
scaling factors for the \etalnu\ ($\gamma\gamma$) decay channel.}
\begin{center}
\begin{tabular}{lccccccc}
\hline\hline
\\[-11pt]
& $p^{\eta,\gamma\gamma}_{1}$ &$p^{\eta,\gamma\gamma}_{2}$ &$p^{\eta,\gamma\gamma}_{3}$ 
&$p^{\eta,\gamma\gamma}_{4}$ &$p^{\eta,\gamma\gamma}_{5}$
&$p^{B\bar{B}}_{1}$&$p^{cont}_{1}$ \\[+1pt] \hline
$p^{\eta,\gamma\gamma}_{1}$ & 100&18&1&-1&9&17&-59\\
$p^{\eta,\gamma\gamma}_{2}$ & 18&100&9&10&13&-17&-23\\
$p^{\eta,\gamma\gamma}_{3}$ & 1&9&100&12&10&-29&5\\
$p^{\eta,\gamma\gamma}_{4}$ & -1&10&12&100&12&-37&11\\
$p^{\eta,\gamma\gamma}_{5}$ & 9&13&10&12&100&-26&-7\\
$p^{B\bar{B}}_{1}$&17&-17&-29&-37&-26&100&-45\\
$p^{cont}_{1}$&-59&-23&5&11&-7&-45&100\\
\hline\hline
\end{tabular}
\end{center}
\end{table*}

\begin{table*}
\caption[]{\label{table6} Correlation matrix (elements in \%) of the fitted 
scaling factors for the \etalnu\ ($\pi^+\pi^-\pi^0$) decay channel.}
\begin{center}
\begin{tabular}{lcc}
\hline\hline
\\[-11pt]
& $p^{\eta,3\pi}_{1}$ &$p^{B\bar{B}}_{1}$ \\[+1pt] \hline
$p^{\eta,3\pi}_{1}$ & 100&-48\\
$p^{B\bar{B}}_{1}$&-48&100\\
\hline\hline
\end{tabular}
\end{center}
\end{table*}

\begin{table*}
\caption[]{\label{table7} Correlation matrix (elements in \%) of the fitted 
scaling factors for the \etalnu\ ($\gamma\gamma$ + $3\pi$) decay channel.}
\begin{center}
\begin{tabular}{lccccccc}
\hline\hline
\\[-11pt]
& $p^{\eta}_{1}$ &$p^{\eta}_{2}$ &$p^{\eta}_{3}$ &$p^{\eta}_{4}$ &$p^{\eta}_{5}$
&$p^{B\bar{B}}_{1}$&$p^{cont}_{1}$ \\[+1pt] \hline
$p^{\eta}_{1}$ & 100&17&1&0&7&18&-57\\
$p^{\eta}_{2}$ & 17&100&5&6&8&-9&-23\\
$p^{\eta}_{3}$ & 1&5&100&8&6&-22&3\\
$p^{\eta}_{4}$ & 0&6&8&100&8&-29&7\\
$p^{\eta}_{5}$ & 7&8&6&8&100&-19&-6\\
$p^{B\bar{B}}_{1}$&18&-9&-22&-29&-19&100&-47\\
$p^{cont}_{1}$&-57&-23&3&7&-6&-47&100\\
\hline\hline
\end{tabular}
\end{center}
\end{table*}

\begin{table*}
\caption[]{\label{table8} Correlation matrix (elements in \%) of the fitted 
scaling factors for the \etaplnu\ ($\gamma\gamma$) decay channel.}
\begin{center}
\begin{tabular}{lcc}
\hline\hline
\\[-11pt]
& $p^{\eta^{\prime}}_{1}$ &$p^{B\bar{B}}_{1}$ \\[+1pt] \hline
$p^{\eta^{\prime}}_{1}$ & 100&-58\\
$p^{B\bar{B}}_{1}$&-58&100\\
\hline\hline
\end{tabular}
\end{center}
\end{table*}

\begin{table*}
\caption[]{\label{picerror} Combined \pilnu\ and \pizlnu\ yields, 
efficiencies (\%), \DBR\ $(10^{-7})$ and their relative uncertainties (\%). The
\DBR\ and efficiency values labeled ``without FSR'' are modified to remove FSR 
effects. This procedure has no significant impact on the \DBR\ values.}
\begin{smaller}
\begin{center}
\begin{tabular}{lcccccccccccccccc}
\hline\hline
\qq\ bins (\gevsq) & 0-2 & 2-4 & 4-6 & 6-8 & 8-10 & 10-12 & 12-14 & 14-16 & 16-18& 18-20& 20-22& 22-26.4& \qq$<$12 & \qq$<$16 & \qq$>$16 & 
Total \\ 
\hline
Fitted yield & 849.9& 1210.4& 1447.3& 1557.9& 1374.9& 1293.0& 959.4& 866.8& 870.2& 672.6& 710.4& 635.3& 7733.2 &9559.4& 2888.5& 12447.9\\
Yield statistical error & 9.9& 6.5& 5.3& 5.4& 6.0& 6.0& 7.5& 8.1& 8.1& 10.5& 10.4& 12.1& 2.6 &3.0& 5.6& 2.9 \\
Unfolded yield & 889.1& 1207.8& 1431.5& 1581.6& 1327.2& 1368.2& 926.5& 887.0& 948.3& 633.9& 699.5& 547.3& 7805.4& 9618.9& 2829.0& 12447.9\\
\hline
$\pi^-\ell\nu$ Efficiency         &6.18& 7.39& 8.52& 8.38& 7.94& 7.00& 6.45& 6.13& 6.34& 6.79& 6.77& 6.06& -& -& -& -\\
$\pi^-\ell\nu$ Eff. (without FSR) &5.93& 7.29& 8.46& 8.46& 7.99& 7.06& 6.49& 6.21& 6.38& 6.88& 6.83& 6.11& -& -& -& -\\
$\pi^0\ell\nu$ Efficiency         &4.00& 5.14& 5.88& 6.03& 5.21& 4.19& 3.22& 2.81& 2.92& 3.44& 4.45& 3.87& -& -& -& -\\
$\pi^0\ell\nu$ Eff. (without FSR) &3.95& 5.15& 5.92& 6.05& 5.24& 4.19& 3.22& 2.80& 2.93& 3.45& 4.41& 3.85& -& -& -& -\\
\hline\hline
\end{tabular}
\end{center}
\end{smaller}
\end{table*}

\begin{table*}
\caption[]{\label{pierror} \pilnu\ yields, efficiencies (\%), \DBR\ 
$(10^{-7})$ and their relative uncertainties (\%). The \DBR\ and efficiency 
values labeled ``without FSR'' are modified to remove FSR effects. This 
procedure has no significant impact on the \DBR\ values.}
\begin{smaller}
\begin{center}
\begin{tabular}{lcccccccccccccccc}
\hline\hline
\qq\ bins (\gevsq) & 0-2 & 2-4 & 4-6 & 6-8 & 8-10 & 10-12 & 12-14 & 14-16 & 16-18& 18-20& 20-22& 22-26.4& \qq$<$12 & \qq$<$16 & \qq$>$16 & 
Total \\ 
\hline
Fitted yield & 630.6& 846.6& 992.7& 1111.7& 1022.7& 966.0& 734.1& 662.3& 687.2& 547.0& 656.8& 439.0& 5570.2&6966.6& 2329.9& 9296.5 \\
Yield statistical error & 12.2& 8.2& 6.5& 6.3& 6.9& 7.1& 8.9& 9.6& 9.7& 12.0& 10.7& 15.9& 3.1 &3.6& 6.5& 3.4 \\
Unfolded yield  & 649.8& 832.6& 964.3& 1133.4& 1003.0& 1021.0& 714.6& 663.8& 731.7& 502.3& 713.0& 367.1& 5604.1& 6982.4& 2314.2& 9296.5 \\
\hline
Efficiency         &6.18& 7.39& 8.52& 8.38& 7.94& 7.00& 6.45& 6.13& 6.34& 6.79& 6.77& 6.06& -& -& -& -\\
Eff. (without FSR) &5.93& 7.29& 8.46& 8.46& 7.99& 7.06& 6.49& 6.21& 6.38& 6.88& 6.83& 6.11& -& -& -& -\\
\hline
\DBR\ & 117.5& 125.9& 126.6& 151.2& 141.3& 163.1& 123.9& 121.1& 129.0& 82.7& 117.8& 67.8& 825.6& 1070.6& 397.3& 1467.9  \\
\DBR\ (without FSR) & 122.5& 127.7& 127.4& 149.8& 140.5& 161.7& 123.1& 119.6& 128.3& 81.6& 116.7& 67.2& 829.6& 1072.3& 393.8& 1466.1 \\
\hline\hline
\end{tabular}
\end{center}
\end{smaller}
\end{table*}

\begin{table*}
\caption[]{\label{pizerror} \pizlnu\ yields, efficiencies (\%), \DBR\ 
$(10^{-7})$ and their relative uncertainties (\%). The \DBR\ and efficiency 
values labeled ``without FSR'' are modified to remove FSR effects. This 
procedure has no significant impact on the \DBR\ values.}
\begin{smaller}
\begin{center}
\begin{tabular}{lccccccccccccccc}
\hline\hline
\qq\ bins (\gevsq) & 0-2 & 2-4 & 4-6 & 6-8 & 8-10 & 10-12 & 12-14 & 14-16 & 16-18& 18-20& 20-26.4& \qq$<$12 & \qq$<$16 & \qq$>$16 & 
Total \\ 
\hline
Fitted yield & 236.3& 386.7& 452.2& 449.2& 351.3& 320.5& 230.8& 189.9& 156.0& 138.8& 292.5& 2196.1 & 2616.8& 587.3& 3204.1 \\
Yield statistical error & 14.5& 9.7& 9.0& 9.3& 11.5& 11.3& 13.3& 15.0& 18.0& 24.9& 27.2& 4.3 & 4.9& 17.0& 5.3 \\
Unfolded yield  & 259.6& 408.5& 453.5& 448.3& 322.9& 339.0& 225.1& 210.0& 175.7& 140.3& 221.4& 2231.7 &2666.7& 537.3& 3204.1 \\
\hline
Efficiency         &4.00& 5.14& 5.88& 6.03& 5.21& 4.19& 3.22& 2.81& 2.92& 3.44& 4.18& -& -& -& -\\
Eff. (without FSR) &3.95& 5.15& 5.92& 6.05& 5.24& 4.19& 3.22& 2.80& 2.93& 3.45& 4.15& -& -& -& -\\
\hline
\DBR\ & 68.0& 83.3& 80.9& 78.0& 65.0& 84.8& 73.3& 78.3& 63.1& 42.8& 55.6& 460.0 & 611.6& 161.5& 773.1  \\
\DBR\ (without FSR) & 68.9& 83.2& 80.3& 77.8& 64.6& 84.9& 73.3& 78.6& 62.9& 42.7& 56.0& 459.7 & 611.6& 161.5& 773.1  \\
\hline\hline
\end{tabular}
\end{center}
\end{smaller}
\end{table*}

\begin{table*}
\caption[]{\label{omegaetacerror} \omegalnu\ and combined \etalnu\ ($3\pi$ and 
$\gamma\gamma$ decay channels) yields, efficiencies(\%), \DBR\ $(10^{-7})$ and 
their relative uncertainties (\%).}
\begin{smaller}
\begin{center}
\begin{tabular}{lccccccp{0.3cm}cccccc}
\hline\hline
Decay mode &  \multicolumn{6}{c}{$\omega\ell^+\nu$ }& & \multicolumn{6}{c}{$\eta\ell^+\nu$ ($3\pi$ and $\gamma\gamma$ combined)}\\
\qq\ bins (\gevsq)      & 0-4  & 4-8  & 8-12 & 12-16 & 16-20.2 & 0-20.2 & & 0-4  & 4-8  & 8-2  & 12-16 & 16-22.4 & 0-22.4 \\ 
\hline
Fitted yield            & 292.6& 567.5& 217.6& 253.5 & 529.7   & 1860.8  & & 231.4& 348.1& 153.1& 93.7 & 41.0   & 867.3   \\
Yield statistical error & 20.5 & 12.1 & 18.5 &  23.4 & 27.6    & 12.5   & & 20.9 &  13.9& 23.8 & 36.2  & 69.6      & 11.6\\
Unfolded yield          & 282.2& 590.8& 267.7&  301.7& 418.5   &   1860.8& & 231.4& 349.8& 155.7& 96.1 & 34.3    & 867.3 \\
\hline
Efficiency              & 2.14 & 2.19 & 0.99 &  1.28 & 1.83    & -     & &2.64  & 3.48 & 1.93 & 1.51    & 1.20      & - \\
\hline
\DBR\                   & 138.1& 283.0& 284.6&  246.7 &  239.8   &   1192.2& & 91.7& 105.4& 84.7& 66.8 & 30.0    & 378.6 \\
\hline
Tracking efficiency     & 6.1  & 2.0  & 2.2  &  5.9  &    2.6 &   2.3  & & 3.1  & 0.2  & 1.2 & 2.4   & 17.3     & 1.1 \\
\hline\hline
\end{tabular}
\end{center}
\end{smaller}
\end{table*}

\begin{table*}
\caption[]{\label{etaerror} \etaetaplnu\  yields, efficiencies(\%), \DBR\ 
$(10^{-7})$ and their relative uncertainties (\%).}
\begin{smaller}
\begin{center}
\begin{tabular}{lcp{0.2cm}cp{0.2cm}cccccc}
\hline\hline
Decay mode & $\eta^{\prime}\ell^+\nu$ & & $\eta\ell^+\nu$ ($3\pi$)& & \multicolumn{6}{c}{$\eta\ell^+\nu$ ($\gamma\gamma$)}\\
\qq\ bins (\gevsq)      & 0-18.7 & & 0-22.4 & &  0-4  &  4-8   &  8-12  &  12-16 &  16-22.4 &  0-22.4  \\ 
\hline
Fitted yield            & 141.1  & &  279.8 & & 192.3 &  186.1 &  105.5 &  49.5 &    36.8 &  570.1 \\
Yield statistical error &  25.6  & &   22.1 & &  22.4 &   21.6 &   27.9 &   50.4 &   71.2 & 15.3 \\
Unfolded yield          &  141.1 & &   279.8& &  201.7&   173.5&   112.6&   46.8&     35.5& 570.1\\
\hline
Efficiency              &   0.61 & &   0.63 & &  1.99 &   2.56 &   1.32 &  1.03    &   0.89 & -  \\
\hline
\DBR\                   &   242.3& &   464.4& &  106.4&   71.2 &   89.7 &   47.5&     42.0& 356.9  \\
\hline
Tracking efficiency     &   4.1  & &   1.0  & &  2.1  &   0.7  &   3.0 &  5.5  &     9.3  & 0.8 \\
\hline\hline
\end{tabular}
\end{center}
\end{smaller}
\end{table*}

\begin{table*}
\caption[]{\label{StatCovPic} Correlation matrix of the partial \bfpiclnuq\ 
statistical uncertainties.}
\begin{center}
\begin{tabular}{lcccccccccccc}
\hline\hline
\qq\ bins (\gevsq) & 0-2 & 2-4 & 4-6 & 6-8 & 8-10 & 10-12 & 12-14 & 14-16 & 16-18& 18-20 & 20-22 & 22-26.4\\ \hline
0-2 & 1.00 & -0.11 & 0.20 & 0.02 & -0.01 & 0.03 & 0.01 & 0.03 & 0.05 & 0.09 & 0.08 & -0.02\\
2-4 & -0.11 & 1.00 & -0.31 & 0.14 & -0.02 & 0.01 & -0.00 & 0.01 & 0.02 & 0.03 & 0.03 & -0.01\\
4-6 & 0.20 & -0.31 & 1.00 & -0.30 & 0.16 & 0.02 & 0.06 & 0.05 & 0.07 & 0.02 & 0.02 & -0.00\\
6-8 & 0.02 & 0.14 & -0.30 & 1.00 & -0.24 & 0.14 & 0.06 & 0.05 & 0.07 & 0.00 & 0.01 & 0.00\\
8-10 & -0.01 & -0.02 & 0.16 & -0.24 & 1.00 & -0.24 & 0.16 & 0.04 & 0.07 & -0.01 & -0.01 & 0.01\\
10-12 & 0.03 & 0.01 & 0.02 & 0.14 & -0.24 & 1.00 & -0.18 & 0.09 & 0.07 & 0.00 & 0.02 & -0.00\\
12-14 & 0.01 & -0.00 & 0.06 & 0.06 & 0.16 & -0.18 & 1.00 & -0.20 & 0.13 & -0.01 & 0.03 & -0.01\\
14-16 & 0.03 & 0.01 & 0.05 & 0.05 & 0.04 & 0.09 & -0.20 & 1.00 & -0.06 & 0.01 & -0.02 & -0.01\\
16-18 & 0.05 & 0.02 & 0.07 & 0.07 & 0.07 & 0.07 & 0.13 & -0.06 & 1.00 & -0.19 & 0.09 & -0.06\\
18-20 & 0.09 & 0.03 & 0.02 & 0.00 & -0.01 & 0.00 & -0.01 & 0.01 & -0.19 & 1.00 & 0.03 & -0.06\\
20-22 & 0.08 & 0.03 & 0.02 & 0.01 & -0.01 & 0.02 & 0.03 & -0.02 & 0.09 & 0.03 & 1.00 & -0.37\\
22-26.4 & -0.02 & -0.01 & -0.00 & 0.00 & 0.01 & -0.00 & -0.01 & -0.01 & -0.06 & -0.06 & -0.37 & 1.00\\
\hline\hline
\end{tabular}
\end{center}
\end{table*}

\begin{table*}
\caption[]{\label{StatCovPi} Correlation matrix of the partial \bfpilnuq\ 
statistical uncertainties.}
\begin{center}
\begin{tabular}{lcccccccccccc}
\hline\hline
\qq\ bins (\gevsq) & 0-2 & 2-4 & 4-6 & 6-8 & 8-10 & 10-12 & 12-14 & 14-16 & 16-18& 18-20 & 20-22 & 22-26.4\\ \hline
0-2 & 1.00 & -0.07 & 0.20 & 0.02 & -0.01 & 0.03 & 0.01 & 0.03 & 0.05 & 0.07 & 0.06 & -0.01\\
2-4 & -0.07 & 1.00 & -0.28 & 0.12 & -0.02 & 0.00 & -0.01 & 0.01 & 0.02 & 0.02 & 0.02 & -0.01\\
4-6 & 0.20 & -0.28 & 1.00 & -0.29 & 0.14 & 0.02 & 0.05 & 0.05 & 0.07 & 0.02 & 0.02 & 0.00\\
6-8 & 0.02 & 0.12 & -0.29 & 1.00 & -0.22 & 0.13 & 0.06 & 0.05 & 0.06 & 0.00 & 0.01 & 0.00\\
8-10 & -0.01 & -0.02 & 0.14 & -0.22 & 1.00 & -0.23 & 0.15 & 0.04 & 0.06 & -0.01 & -0.00 & 0.01\\
10-12 & 0.03 & 0.00 & 0.02 & 0.13 & -0.23 & 1.00 & -0.16 & 0.08 & 0.07 & 0.00 & 0.02 & -0.00\\
12-14 & 0.01 & -0.01 & 0.05 & 0.06 & 0.15 & -0.16 & 1.00 & -0.18 & 0.13 & -0.00 & 0.04 & -0.01\\
14-16 & 0.03 & 0.01 & 0.05 & 0.05 & 0.04 & 0.08 & -0.18 & 1.00 & -0.04 & 0.01 & -0.01 & -0.01\\
16-18 & 0.05 & 0.02 & 0.07 & 0.06 & 0.06 & 0.07 & 0.13 & -0.04 & 1.00 & -0.16 & 0.10 & -0.05\\
18-20 & 0.07 & 0.02 & 0.02 & 0.00 & -0.01 & 0.00 & -0.00 & 0.01 & -0.16 & 1.00 & 0.05 & -0.05\\
20-22 & 0.06 & 0.02 & 0.02 & 0.01 & -0.00 & 0.02 & 0.04 & -0.01 & 0.10 & 0.05 & 1.00 & -0.30\\
22-26.4 & -0.01 & -0.01 & 0.00 & 0.00 & 0.01 & -0.00 & -0.01 & -0.01 & -0.05 & -0.05 & -0.30 & 1.00\\
\hline\hline
\end{tabular}
\end{center}
\end{table*}

\begin{table*}
\caption[]{\label{StatCovPiz} Correlation matrix of the partial \bfpizlnuq\ 
statistical uncertainties.}
\begin{center}
\begin{tabular}{lcccccccccccc}
\hline\hline
\qq\ bins (\gevsq) & 0-2 & 2-4 & 4-6 & 6-8 & 8-10 & 10-12 & 12-14 & 14-16 & 16-18& 18-20 & 20-26.4\\ \hline
0-2 & 1.00 & -0.21 & 0.17 & 0.01 & 0.03 & 0.01 & 0.01 & 0.01 & 0.04 & 0.05 & 0.02\\
2-4 & -0.21 & 1.00 & -0.37 & 0.18 & -0.02 & 0.02 & 0.01 & 0.02 & 0.03 & 0.03 & 0.01\\
4-6 & 0.17 & -0.37 & 1.00 & -0.40 & 0.18 & -0.01 & 0.04 & 0.03 & 0.04 & 0.04 & -0.01\\
6-8 & 0.01 & 0.18 & -0.40 & 1.00 & -0.38 & 0.16 & -0.00 & 0.04 & 0.03 & 0.03 & -0.02\\
8-10 & 0.03 & -0.02 & 0.18 & -0.38 & 1.00 & -0.33 & 0.15 & 0.02 & 0.04 & 0.04 & -0.05\\
10-12 & 0.01 & 0.02 & -0.01 & 0.16 & -0.33 & 1.00 & -0.32 & 0.12 & 0.01 & 0.03 & -0.04\\
12-14 & 0.01 & 0.01 & 0.04 & -0.00 & 0.15 & -0.32 & 1.00 & -0.28 & 0.09 & 0.02 & -0.03\\
14-16 & 0.01 & 0.02 & 0.03 & 0.04 & 0.02 & 0.12 & -0.28 & 1.00 & -0.23 & 0.07 & -0.03\\
16-18 & 0.04 & 0.03 & 0.04 & 0.03 & 0.04 & 0.01 & 0.09 & -0.23 & 1.00 & -0.20 & 0.03\\
18-20 & 0.05 & 0.03 & 0.04 & 0.03 & 0.04 & 0.03 & 0.02 & 0.07 & -0.20 & 1.00 & -0.18\\
20-26.4 & 0.02 & 0.01 & -0.01 & -0.02 & -0.05 & -0.04 & -0.03 & -0.03 & 0.03 & -0.18 & 1.00\\
\hline\hline
\end{tabular}
\end{center}
\end{table*}


\begin{table*}
\caption[]{\label{SystCovPic} Correlation matrix of the partial \bfpiclnuq\ 
systematic uncertainties.}
\begin{center}
\begin{tabular}{lcccccccccccc}
\hline\hline
\qq\ bins (\gevsq) & 0-2 & 2-4 & 4-6 & 6-8 & 8-10 & 10-12 & 12-14 & 14-16 & 16-18& 18-20 & 20-22 & 22-26.4\\ \hline
0-2 & 1.00 & 0.02 & 0.10 & 0.39 & 0.51 & 0.35 & 0.58 & 0.43 & 0.09 & 0.10 & 0.29 & -0.15\\
2-4 & 0.02 & 1.00 & 0.55 & 0.67 & 0.37 & 0.46 & 0.49 & 0.43 & 0.08 & 0.03 & 0.35 & -0.06\\
4-6 & 0.10 & 0.55 & 1.00 & 0.61 & 0.60 & 0.32 & 0.49 & 0.53 & -0.05 & 0.30 & 0.52 & -0.21\\
6-8 & 0.39 & 0.67 & 0.61 & 1.00 & 0.41 & 0.78 & 0.52 & 0.41 & 0.34 & -0.05 & 0.47 & 0.08\\
8-10 & 0.51 & 0.37 & 0.60 & 0.41 & 1.00 & -0.00 & 0.78 & 0.88 & -0.36 & 0.58 & 0.58 & -0.49\\
10-12 & 0.35 & 0.46 & 0.32 & 0.78 & -0.00 & 1.00 & 0.28 & 0.08 & 0.65 & -0.32 & 0.22 & 0.27\\
12-14 & 0.58 & 0.49 & 0.49 & 0.52 & 0.78 & 0.28 & 1.00 & 0.80 & -0.03 & 0.44 & 0.41 & -0.41\\
14-16 & 0.43 & 0.43 & 0.53 & 0.41 & 0.88 & 0.08 & 0.80 & 1.00 & -0.30 & 0.63 & 0.53 & -0.61\\
16-18 & 0.09 & 0.08 & -0.05 & 0.34 & -0.36 & 0.65 & -0.03 & -0.30 & 1.00 & -0.53 & -0.01 & 0.57\\
18-20 & 0.10 & 0.03 & 0.30 & -0.05 & 0.58 & -0.32 & 0.44 & 0.63 & -0.53 & 1.00 & 0.35 & -0.67\\
20-22 & 0.29 & 0.35 & 0.52 & 0.47 & 0.58 & 0.22 & 0.41 & 0.53 & -0.01 & 0.35 & 1.00 & -0.15\\
22-26.4 & -0.15 & -0.06 & -0.21 & 0.08 & -0.49 & 0.27 & -0.41 & -0.61 & 0.57 & -0.67 & -0.15 & 1.00\\
\hline\hline
\end{tabular}
\end{center}
\end{table*}

\begin{table*}
\caption[]{\label{SystCovPi} Correlation matrix of the partial \bfpilnuq\ 
systematic uncertainties.}
\begin{center}
\begin{tabular}{lcccccccccccc}
\hline\hline
\qq\ bins (\gevsq) & 0-2 & 2-4 & 4-6 & 6-8 & 8-10 & 10-12 & 12-14 & 14-16 & 16-18& 18-20 & 20-22 & 22-26.4\\ \hline
0-2 & 1.00 & -0.10 & 0.10 & 0.42 & 0.60 & 0.42 & 0.57 & 0.47 & 0.18 & 0.11 & 0.31 & -0.19\\
2-4 & -0.10 & 1.00 & 0.48 & 0.59 & 0.41 & 0.48 & 0.47 & 0.45 & 0.18 & 0.05 & 0.30 & -0.05\\
4-6 & 0.10 & 0.48 & 1.00 & 0.69 & 0.55 & 0.46 & 0.35 & 0.41 & 0.10 & 0.16 & 0.44 & 0.10\\
6-8 & 0.42 & 0.59 & 0.69 & 1.00 & 0.75 & 0.75 & 0.66 & 0.65 & 0.27 & 0.08 & 0.61 & -0.06\\
8-10 & 0.60 & 0.41 & 0.55 & 0.75 & 1.00 & 0.54 & 0.74 & 0.85 & 0.09 & 0.33 & 0.49 & -0.35\\
10-12 & 0.42 & 0.48 & 0.46 & 0.75 & 0.54 & 1.00 & 0.57 & 0.49 & 0.49 & -0.17 & 0.59 & 0.07\\
12-14 & 0.57 & 0.47 & 0.35 & 0.66 & 0.74 & 0.57 & 1.00 & 0.72 & 0.33 & 0.21 & 0.36 & -0.29\\
14-16 & 0.47 & 0.45 & 0.41 & 0.65 & 0.85 & 0.49 & 0.72 & 1.00 & 0.15 & 0.40 & 0.40 & -0.46\\
16-18 & 0.18 & 0.18 & 0.10 & 0.27 & 0.09 & 0.49 & 0.33 & 0.15 & 1.00 & -0.37 & 0.47 & 0.25\\
18-20 & 0.11 & 0.05 & 0.16 & 0.08 & 0.33 & -0.17 & 0.21 & 0.40 & -0.37 & 1.00 & -0.38 & -0.71\\
20-22 & 0.31 & 0.30 & 0.44 & 0.61 & 0.49 & 0.59 & 0.36 & 0.40 & 0.47 & -0.38 & 1.00 & 0.33\\
22-26.4 & -0.19 & -0.05 & 0.10 & -0.06 & -0.35 & 0.07 & -0.29 & -0.46 & 0.25 & -0.71 & 0.33 & 1.00\\
\hline\hline
\end{tabular}
\end{center}
\end{table*}

\begin{table*}
\caption[]{\label{SystCovPiz} Correlation matrix of the partial \bfpizlnuq\ 
systematic uncertainties.}
\begin{center}
\begin{tabular}{lcccccccccccc}
\hline\hline
\qq\ bins (\gevsq) & 0-2 & 2-4 & 4-6 & 6-8 & 8-10 & 10-12 & 12-14 & 14-16 & 16-18& 18-20 & 20-26.4\\ \hline
0-2 & 1.00 & 0.25 & -0.32 & -0.06 & -0.25 & 0.27 & 0.28 & 0.04 & 0.07 & 0.15 & -0.31\\
2-4 & 0.25 & 1.00 & 0.40 & 0.76 & 0.21 & 0.65 & 0.60 & 0.12 & 0.01 & 0.16 & 0.01\\
4-6 & -0.32 & 0.40 & 1.00 & 0.59 & 0.63 & 0.34 & 0.31 & 0.18 & 0.02 & -0.17 & 0.34\\
6-8 & -0.06 & 0.76 & 0.59 & 1.00 & 0.40 & 0.63 & 0.53 & 0.17 & 0.17 & 0.18 & 0.25\\
8-10 & -0.25 & 0.21 & 0.63 & 0.40 & 1.00 & 0.19 & 0.25 & 0.13 & 0.00 & -0.15 & 0.42\\
10-12 & 0.27 & 0.65 & 0.34 & 0.63 & 0.19 & 1.00 & 0.61 & 0.39 & 0.32 & 0.38 & 0.01\\
12-14 & 0.28 & 0.60 & 0.31 & 0.53 & 0.25 & 0.61 & 1.00 & 0.28 & 0.22 & 0.36 & -0.08\\
14-16 & 0.04 & 0.12 & 0.18 & 0.17 & 0.13 & 0.39 & 0.28 & 1.00 & 0.31 & 0.32 & -0.18\\
16-18 & 0.07 & 0.01 & 0.02 & 0.17 & 0.00 & 0.32 & 0.22 & 0.31 & 1.00 & 0.59 & -0.06\\
18-20 & 0.15 & 0.16 & -0.17 & 0.18 & -0.15 & 0.38 & 0.36 & 0.32 & 0.59 & 1.00 & -0.08\\
20-26.4 & -0.31 & 0.01 & 0.34 & 0.25 & 0.42 & 0.01 & -0.08 & -0.18 & -0.06 & -0.08 & 1.00\\
\hline\hline
\end{tabular}
\end{center}
\end{table*}

\begin{table*}
\caption[]{\label{CovOmega} Correlation matrix of the partial \bfomegalnuq\ 
statistical and systematic uncertainties.}
\begin{center}
\begin{tabular}{lp{0.1cm}cccccp{0.2cm}ccccc}
\hline\hline
 & &\multicolumn{5}{c}{Statistical} & & \multicolumn{5}{c}{Systematic}\\
\qq\ bins (\gevsq)& & 0-4 & 4-8 & 8-12 & 12-16 & 16-20.2 & &0-4 & 4-8 & 8-12 & 12-16 & 16-20.2 \\ \hline
0-4     & & 1.00 & -0.12 & 0.07 & 0.10 & -0.02 & & 1.00 & 0.23& -0.25& -0.20& -0.38 \\
4-8     & & -0.12 & 1.00 & -0.13 & 0.03 & 0.08 & & 0.23 & 1.00& -0.53& -0.62& 0.17 \\
8-12    & & 0.07 & -0.13 & 1.00 & -0.08 & 0.19 & & -0.25 & -0.53& 1.00& 0.82& -0.16 \\
12-16   & & 0.10 & 0.03 & -0.08 & 1.00 & 0.12 & & -0.20 & -0.62& 0.82& 1.00& -0.22 \\
16-20.2 & & -0.02 & 0.08 & 0.19 & 0.12 & 1.00 & & -0.38 & 0.17& -0.16& -0.22& 1.00 \\ \hline\hline 
\end{tabular}
\end{center}
\end{table*}

\begin{table*}
\caption[]{\label{CovEta} Correlation matrix of the partial \bfetalnuq\ 
statistical and systematic uncertainties.}
\begin{center}
\begin{tabular}{lp{0.1cm}cccccp{0.2cm}ccccc}
\hline\hline
 & &\multicolumn{5}{c}{Statistical} & & \multicolumn{5}{c}{Systematic}\\
\qq\ bins (\gevsq)& & 0-4 & 4-8 & 8-12 & 12-16 & 16-22.4 & &0-4 & 4-8 & 8-12 & 12-16 & 16-22.4 \\ \hline
0-4     & & 1.00 & -0.04 & 0.02 & -0.02 & 0.06 & & 1.00 & -0.31& 0.46& 0.23& 0.26 \\
4-8     & & -0.04 & 1.00 & -0.12 & 0.06 & 0.06 & & -0.31 & 1.00& 0.12& -0.17& 0.00 \\
8-12    & & 0.02 & -0.12 & 1.00 & -0.07 & 0.06 & & 0.46 & 0.12& 1.00& 0.45& 0.49 \\
12-16   & & -0.02 & 0.06 & -0.07 & 1.00 & -0.14 & & 0.23 & -0.17& 0.45& 1.00& 0.62 \\
16-22.4 & & 0.06 & 0.06 & 0.06 & -0.14 & 1.00 & & 0.26 & 0.00& 0.49& 0.62& 1.00 \\ \hline\hline 
\end{tabular}
\end{center}
\end{table*}

\begin{thebibliography}{99}
\bibitem{PlusCC}
Charge conjugate decays are implied throughout this paper.
\end{thebibliography}
\end{document}